\documentclass[11pt,reqno]{amsart}
\usepackage{amsfonts}
\usepackage{mathrsfs}
\usepackage{graphicx} 
\usepackage{epsfig}
\usepackage{amsmath, amsthm}
\usepackage{amssymb}

\theoremstyle{plain}
\newtheorem{thm}{Theorem}[section]
\newtheorem{prop}[thm]{Proposition}
\newtheorem{lem}[thm]{Lemma}
\newtheorem{cor}[thm]{Corollary}

\theoremstyle{definition}

\newtheorem{rem}[thm]{Remark}
\newtheorem{defn}[thm]{Definition}
\newtheorem{eg}[thm]{Example}
\newtheorem{subtitle}[thm]{}
\newtheorem{ex}{Exercise}[section]
\numberwithin{equation}{section}

\def\a{\alpha}

\def\D{\triangle}

\def\g{\gamma}
\def\G{\Gamma}

\def\l{\lambda}
\def\L{\Lambda}
\def\n{\vert\/}

\def\cb{{\mathcal{B}}}

\def\cg{{\mathcal{G}}}

\def\cl{{\mathcal{L}}}

\def\cn{{\mathcal{N}}}
\def\co{{\mathcal{O}}}

\def\ct{{\mathcal{T}}}

\def\cw{{\mathcal{W}}}

\def\n{|\/ }
\def\tr{{\rm tr}}
\def\bs{\bigskip}
\def\ms{\medskip}

\def\ni{\noindent}
\def\ti{\tilde}
\def\p{\partial}

\def\Im{{\rm Im\/}}
\def\I{{\rm I\/}}

\def\diag{{\rm diag}}

\def\Ad{{\rm Ad}}

\def\C{\mathbb{C}}

\def\R{\mathbb{R} }
\def\Z{\mathbb{Z}}

\newcommand{\beq}{\begin{equation}}
\newcommand{\eeq}{\end{equation}}
\newcommand{\beg}{\begin{eg}}
\newcommand{\eeg}{\end{eg}}
\newcommand{\bthm}{\begin{thm}}
\newcommand{\ethm}{\end{thm}}
\newcommand{\bprop}{\begin{prop}}
\newcommand{\eprop}{\end{prop}}
\newcommand{\bcor}{\begin{cor}}
\newcommand{\ecor}{\end{cor}}
\newcommand{\blem}{\begin{lem}}
\newcommand{\elem}{\end{lem}}
\newcommand{\bca}{\begin{cases}}
\newcommand{\eca}{\end{cases}}
\newcommand{\brem}{\begin{rem}}
\newcommand{\erem}{\end{rem}}
\newcommand{\bpm}{\begin{pmatrix}}
\newcommand{\epm}{\end{pmatrix}}
\newcommand{\bbm}{\begin{bmatrix}}
\newcommand{\ebm}{\end{bmatrix}}
\newcommand{\bvm}{\begin{vmatrix}}
\newcommand{\evm}{\end{vmatrix}}
\newcommand{\bdefn}{\begin{defn}}
\newcommand{\edefn}{\end{defn}}
\newcommand{\bsub}{\begin{subtitle}}
\newcommand{\esub}{\end{subtitle}}
\newcommand{\bex}{\begin{ex}}
\newcommand{\eex}{\end{ex}}
\newcommand{\ben}{\begin{enumerate}}
\newcommand{\een}{\end{enumerate}}

\date{}

\def\calB{\mathcal{B}}

\def\calD{\mathcal{D}}

\def\calG{\mathcal{G}}

\def\calL{\mathcal{L}}

\def\calN{\mathcal{N}}
\def\calO{\mathcal{O}}
\def\R{\mathbb{R}}

\def\calT{\mathcal{T}}
\def\calW{\mathcal{W}}

\def\R{\mathbb{R}}

\def\C{\mathbb{C}}

\def\Z{\mathbb{Z}}

\def\diag{{\rm diag \/ }}
\def\det{{\rm det \/ }}

\def\n{\, |\,}

\def\mod{{\rm mod\,}}
\def\bc{\bf c}

\def\rd{{\rm \/ d\/}}
\def\bu{\bullet}

\def\sech{{\rm sech\/}}

\def\an1{A^{(1)}_{n-1}}
\def\Ker{{\rm Ker\/}}

\def\bh{\backslash}
\def\r0{\R^n \backslash \{0\}}

\def\bc{{\bf c}}

\begin{document}

\title[BT for GD$_n$ Flow]
{B\"acklund transformations for Gelfand-Dickey flows, revisited} \today

\author{Chuu-Lian Terng$^\dag$}\thanks{$^\dag$Research supported
in  part by NSF Grant DMS-1109342}
\address{Department of Mathematics\\
University of California at Irvine, Irvine, CA 92697-3875.  Email: cterng@math.uci.edu}
\author{Zhiwei Wu$^*$}\thanks{$^*$Research supported in part by NSF of China under Grant No. 11401327\/}
\address{Department of Mathematics\\ Ningbo University\\ Ningbo, Zhejiang, 315211, China. Email: wuzhiwei@nbu.edu.cn}

\maketitle

\section{Introduction} 

The $j$-th flow of GD$_n$-hierarchy (cf. \cite{Dic03}) is the following evolution equation 
\beq\label{ri}
\frac{\p L}{\p t_j}=[(L^{\frac{j}{n}})_+, L]
\eeq
on the space 
$$\calD_n=\{L= \p^n-\sum_{i=1}^{n-1} u_i \p^{i-1}\n u_i\in C^\infty(\R, \C), 1\leq i\leq n-1\},$$
where $(L^{\frac{j}{n}})_+$ is the differential operator component of the pseudo-differential operator $L^{\frac{j}{n}}$, $\p=\p_x$, and $j\not\equiv 0$ ($\mod n$).

The $\an1$-KdV hierarchy constructed by Drinfeld-Sokolov in \cite{DS84} is the GD$_n$-hierarchy with matrix valued Lax pairs.   The $n\times n$ KdV hierarchy constructed by the first author and Uhlenbeck in \cite{TUa} is equivalent to the GD$_n$-hierarchy.   The $n\times n$ mKdV hierarchy given in  \cite{DS84}  and the KW-hierarchy in \cite{KW81} are equivalent. Moreover, the GD$_n$ and the $n\times n$ mKdV hierarchies are related by the Miura transform.  

B\"acklund transformations (BTs) for the KdV (i.e., GD$_2$) hierarchy constructed from a system of compatible differential equations are well-known (cf. \cite{A78}, \cite{WE}). 

Adler in \cite{A81} used the Miura transform to construct a BT for \eqref{ri} as follows: Suppose $L=\p^n-\sum_{i=1}^{n-1} u_i \p^{i-1}$ is a solution of the $j$-th GD$_n$ flow \eqref{ri}, and $L$ is factored as a product of first order operators
$$L = (\p-q_n) \cdots (\p- q_1)$$
such that $q=\sum_{i=1}^{n-1} q_i e_{ii}$ is a solution of the $n \times n$ mKdV flow. 
Let $\ti L$ be defined by $$\ti L:= (\p- q_{n-1}) \cdots (\p-q_1) (\p- q_n).$$
Then $\ti L$ is a solution of \eqref{ri}. 
Rational solutions are obtained by applying Adler's BT to the vacuum solution $L=\p^n$ (cf. \cite{A81}). 

Note that operators $\ti L$ and $L$ in Adler's result are related by $$\ti L= (\p-q_n)^{-1}L (\p-q_n).$$  
Given a solution $L$ of \eqref{ri}, it is not difficult to see that $\ti L=(\p+h)^{-1}L(\p+h)$ is a solution of \eqref{ri} if and only if $h$ satisfies a system of ordinary differential equations in $x$ and $t$ variables. One goal of this paper is to find all solutions of this system for $h$. To achieve this, we need to use B\"acklund transformations for the $n\times n$ KdV hierarchy constructed in \cite{TUa} and the equivalence of the $n\times n$ KdV and the GD$_n$ hierarchies. 

  Assume $L=\p^n-\sum_{i=1}^{n-1} u_i \p^{(i-1)}$ is a solution of \eqref{ri}. We prove the following results:  
\ben
\item[$\bu$] There exists a system for $h:\R^2\to \C$,
\beq\label{btji} \bca h_x^{(n)}= r_n(u,h),\\ h_t= \eta_{n,j}(u,h),\eca
\eeq
 for some differential polynomials $r_n(u,h)$ and $\eta_{n,j}(u,h)$ such that $\ti L=(\p+h)^{-1}L (\p+h)$ is  a solution of \eqref{ri} if and only if $h$ is a solution of \eqref{btji}. 
\item[$\bu$] There exists a differential polynomial $\xi_n(u,h)$ such that $r_n(u,h)= (\xi_n(u,h))_x$.
\item[$\bu$] If $h$ is a solution of \eqref{btji} then there exists a constant $k\in \C$ such that $h$ satisfies
\beq\label{btni}({\rm BT})_{u,k} \,\, \bca h_x^{(n-1)} =\xi_n(u,h)- k,\\ h_t= \eta_{n,j}(u,h).\eca
\eeq
\item[$\bu$] Systems \eqref{btji} and \eqref{btni} are solvable for $h$.
\item[$\bu$]  $\ti L= (\p+h)^{-1}L (\p+h)$
is a solution of \eqref{ri} if and only if there exists a constant $k\in \C$ such that $h$ is a solution of (BT)$_{u,k}$ (we use $h \ast L$ to denote $\ti L$ and call {\it $L \mapsto h \ast L$ a BT of $L$ with parameter $k$\/}).
\item[$\bu$] There exist differential polynomials $p_{j,1}(u,\l), \ldots, p_{j,n}(u,\l)$ such that the following linear system for $\phi:\R^2\to \C$,
\beq\label{th}
\bca L\phi= \l\phi, \\ \phi_t= \sum_{i=1}^n p_{j,i}(u,\l) \phi_x^{(i-1)},\eca
\eeq
is solvable for all parameter $\l\in \C$, where $p_{j,i}(u,\l)$ has degree $[\frac{j}{n}]$ in $\l$.  Moreover, let $\phi_1, \ldots, \phi_{n-1}$ be linearly independent solutions of \eqref{th} with $\l=k$, and  $W(\phi_1, \ldots, \phi_{n-1})$  the Wronskian of $\phi_1, \ldots, \phi_{n-1}$. Then $$h=(\ln W(\phi_1, \ldots, \phi_{n-1}))_x$$ 
is a solution of (BT)$_{u,k}$ \eqref{btni} and $\ti L= (\p+h)^{-1}L (\p+h)$ is a solution of \eqref{ri}. 
In fact, this construction gives all solutions of \eqref{btni}.
\een

We give algorithms to compute differential polynomials $r_n, \eta_{n,j}$, $\xi_n$, and $p_{j,i}$'s.    For example, the second GD$_3$ flow for $L=\p^3-u_2 \p- u_1$ is \eqref{jt}, \eqref{btji} is \eqref{bt3}, (BT)$_{u,k}$ is \eqref{btn3}, and if $h$ is a solution of \eqref{btn3} then $\ti L= (\p+h)^{-1}L(\p+h)= \p^3- \ti u_2 \p -\ti u_1$ is a new solution of \eqref{jt}, where $\ti u_i$ is given by the formula \eqref{ct}.  System \eqref{th} is \eqref{th3}.

Note that given a solution of \eqref{ri} and a constant $k$, system \eqref{btni} gives rise to $(n-1)$-parameter family of new solutions of \eqref{ri}. When we apply BT with constant $k=0$ and $k\not=0$ to the vacuum solution $L=\p^n$ we obtain explicit rational solutions and soliton solutions respectively.  

We also obtain the following results:
\ben
\item[(a)] We give a  permutability formula for our BTs of the GD$_n$ flows.
\item[(b)]  Adler's BT is our BT with parameter $k=0$.
\item[(c)] We construct a simple and natural action of the order $n$ cyclic group $\Z_n$ on the space of solutions of the KW-hierarchy and show that this $\Z_n$-action gives Adler's BT for the GD$_n$-hierarchy under the Miura transform. 
   \een

This paper is organized as follows: We set up notations and review the constructions and relations of various KdV-type hierarchies in section \ref{km}, construct BTs for the $\an1$-KdV hierarchy in section \ref{btb},  and give a Permutability formula for these BTs and a relation between BTs and scaling transform in section \ref{btp}. We use BTs to construct explicit rational and soliton solutions for the $j$-th $\an1$-KdV flow in section \ref{ns}, and give a natural $\Z_n$-action on the KW-hierarchy and show that Adler's BT arises from this action in section \ref{pia}.   In the last section, we construct BTs for the GD$_n$-hierarchy and explain the relation between our BTs and Adler's BTs.

\bs

\section{Various KdV type hierarchies}\label{km}

In this section, we set up notations and give a brief review of the constructions of the $\an1$-KdV, the $n\times n$ mKdV, the $n\times n$ KdV, and the KW-hierarchies.   

Let $\calB^+_n$, $\cb^-_n$, $\calT_n$, $\calN_n^+$, and $\calN_n^-$ denote the sub-algebras of upper triangular, lower triangular, diagonal,  strictly upper triangular, and strictly lower triangular matrices in $sl(n,\C)$ respectively, and $N_n^+$ the subgroup of upper triangular matrices in $SL(n,\C)$ with $1$ on the diagonal entries.  Let
\begin{align*}
&\calL=\left\{\xi(\l)= \sum_{i\leq m_0} \xi_i \l^i \, \bigg|\, \xi_i\in sl(n,\C), m_0 \,\, {\rm an\,\, integer\,}\right\},\\
& \calL_+=\{\xi(\l)=\sum_{i\geq 0} \xi_i \l^i\in \calL\},\\
&\calL_-=\{\xi(\l)= \sum_{i<0}\xi_i\l^i\in \calL\}.
\end{align*}
Note that $\calL_+, \calL_-$ are Lie subalgebras of $\calL$ and $\calL= \calL_+\oplus \calL_-$ as a direct sum of linear subspaces (such pair $(\calL_+, \calL_-)$ is called a {\it splitting\/} of $\calL$). Given $\xi=\sum_i \xi_i \l^i\in \calL$,  we will use $\xi_\pm$ to denote the projection of $\xi$ onto $\cl_\pm$ w.r.t. $\cl=\cl_+ \oplus \cl_-$, i.e.,
$$\xi_+=\sum_{i\geq 0} \xi_i\l^i, \quad \xi_-=\sum_{i<0} \xi_i \l^i.$$

\ms
\bsub {\bf The $\an 1$-KdV hierarchy \cite{DS84}}

Let 
\beq\label{kt} 
J= e_{1n}\l + b, \quad  b=\sum_{i=1}^{n-1} e_{i+1, i}.
\eeq
Given $u\in C^\infty(\R, \calB_n^+)$, a direct computation shows (cf.  \cite{TUb}) that there exists a unique 
$Q(u, \l)\in  \calL$ such that $Q(0,\l)= J$ and
\beq\label{kta}
\bca[\p_x+J+u, Q(u,\l)]=0,\\
Q(u,\l) \,\, {\rm is\, conjugate\, to\,\,} J.\eca
\eeq 
Write
\beq\label{ktb}
Q^j(u,\l)=\sum_{i\leq [\frac{j}{n}]+1} Q_{j, i}(u) \l^i.
\eeq
Let 
\beq\label{ua}
V_n=\oplus_{i=1}^{n-1}\C e_{in}.
\eeq
It was proved in \cite{DS84} that given $u\in C^\infty(\R, V_n)$ and $j\not\equiv 0$ $(\mod n)$, there exists a unique $\calN_n^+$-valued differential polynomial $\eta_j(u)$ such that
$$[\p_x+b+u, Q_{j,0}(u)+ \eta_j(u)]\,\in \, C^\infty(\R, V_n).$$ 

Let 
\beq\label{pp}
Z_j(u,\l)= (Q^j(u,\l))_+ +\eta_j(u) =\sum_{i=0}^{[\frac{j}{n}]+1} Z_{j,i}(u)\l^i.
\eeq
So we have 
$$ Z_{j,i}(u)= \bca Q_{j,i}(u), & i>0,\\ Q_{j,0}(u)+\eta_j(u), & i=0. \eca$$
The {\it $j$-th flow in the $\an1$-KdV hierarchy\/} is the following evolution equation on $C^\infty(\R, V_n)$, 
\beq\label{dga}
u_{t_j}= [\p_x+ b+ u, Z_{j,0}(u)],
\eeq
where
$b=\sum_{i=1}^{n-1} e_{i+1, i}$.  Moreover, it was proved in \cite{DS84} that $u:\R^2\to V_n$ is a solution of \eqref{dga} if and only if 
\beq\label{laxj}
[\p_x+ J+ u, \p_{t_j}+ Z_j(u,\l)]=0
\eeq
for all parameter $\l\in \C$. 
We call \eqref{laxj} the {\it Lax pair of the solution $u$ of the $j$-th $\an1$-KdV flow\/}. 
\esub

\ms
\bdefn {\bf Frame}

Let $\calO$ be a connected open subset of $\C$. A map $E:\R^2\times \calO\to GL(n,\C)$ is called a {\it frame\/} of the solution $u$ of the $j$-th  $\an1$-KdV flow if $E(x,t,\l)$ is holomorphic for $\l\in \calO$ and  satisfies
$$
\bca E^{-1}E_x= J+ u,\\ E^{-1}E_t= Z_j(u,\l),\eca
$$
where 
$J$ is defined by \eqref{kt}. In other words, $E(\cdot, \cdot, \l)$ is a parallel frame for the Lax pair \eqref{laxj} of $u$. When $u$ is a real solution, we also require $\overline{\calO}= \calO$ and $E$ satisfies  the following reality condition,
$$
\overline{E(x,t,\bar\l)}= E(x,t,\l).
$$
\edefn

Since $Q(0,\l)=J$, $Z_j(0,\l)=J^j$. So $u(x,t)=0$ is a solution (the {\it vacuum solution\/}) of \eqref{dga} and 
$E(x,t,\l)= \exp(xJ+ t J^j)$ is the frame of $u=0$ satisfying $E(0,0,\l)=\I_n$ (the {\it vacuum frame\/}).

The following Theorem was proved by 
Drinfeld-Sokolov.

\bthm\label{so} (\cite{DS84})  $u=\sum_{i=1}^{n-1} u_i e_{in}$ is a solution of the $j$-th $\an1$-KdV flow \eqref{dga}  if and only if $L_u= \p^n-\sum_{i=1}^{n-1} u_i \p^{i-1}$ is a solution of the $j$-th GD$_n$ flow.
\ethm

 Hence the GD$_n$ and the $\an1$-KdV hierarchies are the same, and \eqref{laxj} is a matrix valued Lax pair for the GD$_n$-hierarchy. 

\bsub {\bf The $n \times n$ mKdV hierarchy} (\cite{DS84})

The Drinfeld-Sokolov {\it $n\times n$ mKdV hierarchy\/} is constructed from the following splitting of the loop algebra $\calL$,
\begin{align*}
&\calW_+=\{\xi(\l)=\xi_0+\sum_{i>0} \xi_i\l^i\in \calL\n \xi_0\in \calB_n^-\},\\
&\calW_-=\{\xi(\l)=\xi_0+\sum_{i<0}\xi_i \l^i\in \calL\n \xi_0\in \calN_n^+\}.
\end{align*}
Let $\pi_{bn}$ denote the projection of $sl(n,\C)$ onto $\calB_n^-$ with respect to $sl(n,\C)=\calB_n^-\oplus\calN_n^+$, and let $\pi_+$ denote the projection of $\calL$ onto $\calW_+$ with respect to the direct sum $\calL=\calW_+\oplus \calW_-$ of linear subspaces. Then
\beq\label{io}
\pi_+\left(\sum_{i\leq m_0} \xi_i\l^i\right)= \sum_{i>0} \xi_i \l^ i + \pi_{bn}(\xi_0).
\eeq

It was proved in \cite{DS84} that if $q\in C^\infty(\R, \calT_n)$, then
$[\p_x+ b+ q, \pi_{bn}(Q_{j,0}(q))]$ lies in $C^\infty(\R, \calT_n)$.  The {\it  $j$-th $n\times n$ mKdV flow\/} is the following flow on $C^\infty(\R, \calT_n)$,
\beq\label{im}
q_{t_j} = [\p_x+ b+ q, \pi_{bn}(Q_{j,0}(q))],
\eeq
where $Q_{j,0}(u)$ is defined by \eqref{ktb}. 
Moreover, $q:\R^2\to \calT_n$ is a solution of \eqref{im} if and only if 
$$
[\p_x+J+ q, \p_{t_j} + \pi_{+}(Q^j(q, \l))]=0
$$
for all $\l \in \C$, where $\pi_+$ is the projection of $\cl$ onto $\cw_+$ defined by \eqref{io} and $Q(q,\l)$ is defined by \eqref{kta}. 
\esub

\bsub {\bf The $n \times n$ KdV hierarchy} (\cite{TUa}, \cite{TUb})

The {\it $n\times n$ KdV hierarchy\/} is constructed from an unusual splitting of $\calL$.  Let $\a=e^{\frac{2 \pi i}{n}}$, and 
\beq\label{iw}
\L= \sum_{k=1}^{n-1} \frac{1-\a^k}{1-\a} e_{k, k+1}.
\eeq
Then $\{\L^i b^{n-1}\L^j\n 0\leq i, j\leq n-1\}$ is a basis of $sl(n,\C)$ and $\{\L^ib^{n-1}\L^j\n i+j<n-1\}$ is a basis of $\calN_n^-$.  Let $B: sl(n, \C) \rightarrow \calN_n^+$ be the linear map defined by
$$B(\L^i b^{n-1}\L^j)= \L^i b^t \L^j.$$
Note that $\Ker(B)=\calB_n^+$.  It was proved in \cite{TUb} that
$$\calL_-^B=\{\xi(\l)=B(\xi_{-1})+ \sum_{i<0} \xi_i\l^i\in \calL\}$$ 
is a subalgebra of $\calL$, and 
$$\calL=\calL_+\oplus \calL_-^B$$
is a direct sum of linear subspaces.  Let $\pi_B$ denote the projection of $\calL$ onto $\calL_+$ with respect to the direct sum $\calL=\calL_+\oplus \calL_-^B$. The {\it $j$-th $n\times n$ KdV flow\/} is the following evolution equation on $C^\infty(\R, \oplus_{i=1}^{n-1} \C \L^i)$,
\beq\label{in}
\xi_{t_j}= [\p_x+ b+ \xi, Q_{j,0}(\xi)-B(Q_{j,-1}(\xi))],
\eeq
where $Q_{j,i}(\xi)$ is defined by \eqref{ktb}.
Moreover, $\xi:\R^2\to \oplus_{i=1}^n \C\L^i$ is a solution of \eqref{in} if and only if
\beq\label{ina}
[\p_x+J+\xi, \p_{t_j}+ \pi_B(Q^j(\xi,\l))]=0
\eeq
for all parameter $\l\in \C$, where $Q(\xi,\l)$ is defined by \eqref{kta}.

The $n \times n$ KdV hierarchy can be also constructed from the following equivalent splitting: 
Let $f_i=(1+\a+\cdots+\a^{i-1})^{-1}\L^i$, where $\a=e^{\frac{2 \pi i}{n}}$ and $\L$ is as in \eqref{iw}. Let $\phi_n(z)=I+\sum_{i=1}^{n-1}f_iz^i$,  $\l=z^n$, and $T: \cl \rightarrow \cl$ the algebra homomorphism defined by:
\beq\label{su}
T(A(z))=\phi_n(z)A(z^n)\phi_n(z)^{-1}.
\eeq
Let $\cl_T=T(\cl)$. It was proved in \cite{TUb} that 
$$(\cl_T)_+:= T(\cl_+)=\cl_T\cap \cl_+,  \quad (\cl_T)_-:= T(\cl_-^{B})=\cl_T\cap \cl_-,$$ 
and 
$$T(e_{1n}z^n+b)=az+b,$$ where  $a=\diag(1, \a, \cdots, \a^{n-1})$. Hence the Lax pair \eqref{ina} becomes 
\beq\label{inb}
[\p_x+az+b+\xi, \p_{t_j}+(\ti Q^j(\xi, z))_+]=0, 
\eeq
where $\ti Q(\xi, z)= \phi_n(z)Q(\xi, z^n)\phi_n(z)^{-1}$.
We call \eqref{ina} and \eqref{inb} the {\it Lax pair for $\xi$ in $\l$-guage and $z$-gauge\/} respectively, where $\l= z^n$.  
\esub

We need the following Proposition (proved in \cite{DS84}) to state the equivalence of the $n\times n$ KdV and the $\an1$-KdV hierarchies.

\bprop\label{iz} (\cite{DS84})
Given $v\in C^\infty(\R, \calB_n^+)$, then there exist a unique $N_n^+$-valued differential polynomial $\D$ and $V_n$-valued differential polynomial $u$ in $v$ such that 
\beq\label{oq}
\D (\p_x+ J+ v)\D^{-1}= \p_x+ J+ u,
\eeq
where $V_n$ is defined by \eqref{ua}.
\eprop

\bdefn\label{sm}  Let  $\Psi:C^\infty(\R, \calB_n^+)\to C^\infty(\R, V_n)$ and $\G: C^\infty(\R, \calB_n^+)\to C^\infty(\R, N_n^+)$ be the maps defined by
$$\Psi(v)= u, \quad \G(v)= \D$$
respectively, 
where $v$, $u$, and $\D$ are related by \eqref{oq}.  
\edefn

Let
$$
\calG_j=\bca  \oplus_{i=1}^{n-j} \C e_{i, i+j}, & j\geq 0, \\ \oplus_{i=1-j}^n \C e_{i, i+j}, & j <0. \eca
$$

\bthm\label{ix} (\cite{TUc}) Let $\xi$ be a solution of the $j$-th $n\times n$ KdV flow \eqref{in}, $u(\cdot, t)= \Psi(\xi(\cdot, t))$, and $\D(\cdot, t)= \G(\xi(\cdot, t))$, where $\Psi$ and $\G$ are the maps defined in Definition \ref{sm}.   
 Then we have the following results:
  \ben 
  \item[(a)] $u$ is a solution of the $j$-th $\an1$-KdV flow \eqref{dga}.
  \item[(b)] The map $\Psi$ maps the space of solutions of \eqref{in} bijectively to the space of solutions of \eqref{dga}. 
 \item[(c)] Write $\D= \I_n+ \sum_{i=1}^{n-1} \D_i$ with $\D_i\in \calG_i$, then $\D_1=0$. 
 \item[(d)] If $F$ is a frame of a solution $\xi$ of \eqref{in}, then $E= F\D^{-1}$ is a frame of the solution $u(\cdot, t)$ of the $j$-th $\an1$-KdV flow \eqref{dga}. 
 \een
\ethm

\ms
\bsub {\bf The KW-hierarchy} (\cite{KW81})

 Let 
\begin{align*}
&\tau=e_{21}+e_{32}+\cdots+e_{n, n-1}+e_{1n}, \\
& \a=e^{\frac{2 \pi i}{n}},\\
& a=\diag(1, \a, \cdots, \a^{n-1}).
\end{align*}
The KW-hierarchy is obtained from the splitting $\cl^{KW}_{\pm}$ of the algebra $\cl^{KW}$, where
\begin{align*}
& \cl^{KW}= \{A \in \cl \n A(\a^{-1}z)=\tau A(z) \tau ^{-1}\}, \\
& \cl^{KW}_\pm=\cl^{KW} \cap \cl_\pm.
\end{align*}
Given $v=(v_1, \cdots, v_{n-1})$, let 
\beq\label{qb}
P(v)=\sum_{i=1}^{n-1}v_i\tau^i.
\eeq
Let $\hat Q(v, z)=\sum_{i \leq 1} \hat Q_{1, i}(v)z^i$ be the unique solution of 
$$[\p_x+az+P(v), \hat Q(v, z)]=0$$
satisfying $\hat Q(0, z)=az$ and $\hat Q(v, z)$ is conjugate to $az$. Write 
$$\hat Q^j(v, z)=\sum_{i \leq j}\hat Q_{j, i}(v)z^i.$$
The {\it $j$-th flow in the KW-hierarchy} is the following evolution equation on $C^{\infty}(\R, \oplus_{i=1}^{n-1}\C \tau^i)$,
\beq\label{iq}
\frac{\p P(v)}{\p t_j}=\sum_{i=1}^{n-1} (v_i)_{t_j} \tau^i=[\p_x+P(v), \hat Q_{j, 0}(v)]. 
\eeq
Moreover, $v$ is a solution of \eqref{iq} if and only if 
$$
[\p_x+az+P(v), \p_{t_j}+\left(\hat Q^j(v, z)\right)_+]=0.
$$
\esub

Next we review  relations between the KW, the $n\times n$ mKdV, and the $\an1$-KdV hierarchies. 

\bthm\label{iva} (\cite{TUc})
Let $\Phi:\R^{1\times (n-1)}\to \ct_n$ be the linear isomorphism defined by $\Phi(v_1, \ldots, v_{n-1})= q= \diag(q_1, \ldots, q_n)$, where $q_i=\sum_{k=1}^{n-1}\a^{k(i-1)}v_k$ for $1 \leq i \leq n$ and $\a= \exp(2\pi i/n)$. 
Then we have the following.
\ben
\item[(i)] $v$ is a solution of the $j$-th KW flow \eqref{iq} if and only if $q=\Phi(v)$ is a solution of the $j$-th $n\times n$ mKdV flow \eqref{im}.
\item[(ii)] Let $V(z)=((\a^{(i-1)}z)^{j-1}), \l=z^n$ and $J=e_{1n}\l+b$ as in \eqref{kt}. Then
$$V(z)^{-1} (\p_x+ a z+ P(v))V(z)= \p_x+ J+ q,$$
where $P(v)=\sum_{i=1}^{n-1} v_i \tau^i$ is as defined by \eqref{qb}. 
\een
\ethm

\bthm\label{ivb} (\cite{DS84}) 
Given $q=\sum_{i=1}^nq_ie_{ii}:\R^2\to sl(n,\C)$, let $u_1, \ldots, u_{n-1}$ be defined by 
$$(\p - q_n) \cdots (\p - q_1)= \p^{(n)} -\sum_{i=1}^{n-1} u_i \p^{i-1}.$$
Then we have the following.
\ben
\item[(i)] if $q$ is a solution of the $j$-th $n\times n$ mKdV- flow \eqref{im}, then $ u=\sum_{i=1}^{n-1}  u_i e_{in}$ is a solution of the $j$-th $\an1$-KdV flow \eqref{dga} (this is the {\it Miura transform}). 
\item[(ii)] Let $\G$ and $\Psi$ be the maps defined in Definition \ref{sm}, $\D(\cdot, t)= \G(q(\cdot, t))$, and  $K$ a frame of the solution $q$ of \eqref{im}.  Then $u(\cdot, t)= \Psi(q(\cdot, t))$ and $E= K\D^{-1}$ is a frame of the solution $u$ of \eqref{dga}.  
\een
\ethm

\bs
\section{B\"{a}cklund transformation for $\an1$-KdV}\label{btb}

In this section, we prove the following results:
\ben
\item[(i)] Assume that $E$ is a frame of a solution $u$ of the $j$-th $\an1$-KdV flow \eqref{dga}, and $f$ is of the form $J+ h\I_n+ N$ for some complex valued function $h$ and an $\calN_n^+$-valued map $N$ such that $\ti E= Ef^{-1}$ is also a frame of a solution of \eqref{dga}. Then the entries of $N$ are differential polynomials of $u$ and $h$, and $h$ satisfies the system \eqref{btji}.
\item[(ii)]  If $u$ is a solution of \eqref{dga}, then system \eqref{btji} is solvable and a solution $h$ gives rise to a new solution of \eqref{dga}.
\item[(iii)] All solutions of \eqref{btji} can be constructed from frames of the Lax pair of a solution $u$ of \eqref{dga}.  
\item[(iv)] $h$ is a solution of \eqref{btji} if and only if there exists a constant $k\in\C$ such that $h$ is a solution of \eqref{btni}.
\een 

 B\"acklund transformations for the $n\times n$ KdV hierarchy  is constructed in \cite{TUa} from a loop group factorization with respect to the splitting in $z$-gauge. We use the map $T$ defined by \eqref{su} to state this result in $\l$-gauge as follows:

\bthm\label{or} (\cite{TUa}) Let $j\not\equiv 0$ $ (\mod n)$, and $\L$ defined by \eqref{iw}. Then there exist differential polynomials $A_j(\xi, Y)$ and $B_j(\xi, Y)$ such that  if $F$ is a frame of a solution $\xi$ of the $j$-th $n \times n$ KdV flow \eqref{in}  then $\ti F=F(J+Y)^{-1}$ is a frame of a solution solution $\ti \xi$ of \eqref{in} if and only if $Y$ satisfies the following first order system,
\beq\label{os}
\bca Y_x= A_j(\xi, Y),\\
Y_t= B_j(\xi,Y),\eca
\eeq
 where $J=e_{1n}\l+ b$ is as in \eqref{kt} and $Y=\sum_{i=0}^{n-1}y_i\L^i$.  Moreover, if $\xi$ is a solution of \eqref{in}, then we have the following:
\ben
\item[(i)] System \eqref{os} is solvable for $Y\in C^\infty(\R, \oplus_{i=0}^{n-1} \C \L^i)$.
\item[(ii)] If $Y$ is a solution of \eqref{os}, then there exists a constant $k\in \C$ such that $\det(J+Y)=(-1)^{n-1}(\l-k)$.
\item[(iii)] Given constants $k\in \C$ and $\bc_0\in \C^n\bh 0$, let
$$\zeta(x,t)=(\zeta_1, \cdots, \zeta_n)^t= F(x,t,k)^{-1}(\bc_0),$$ Then there is a unique solution $Y=\sum_{i=0}^{n-1}y_i\L^i$ of \eqref{os} satisfying $y_0=-\frac{\zeta_{n-1}}{\zeta_n}$ and $(ke_{1n}+b+Y)\zeta=0$. 
\item[(iv)] All solutions of \eqref{os} can be constructed as in (iii). 
\een
\ethm

Note that if $\xi$ is a solution of \eqref{in} and $Y$ is a solution of \eqref{os}, then
$$
\ti \xi=(J+Y)(J+\xi)(J+Y)^{-1}-Y_x(J+Y)^{-1}-J
$$
is again a solution of \eqref{in}.  By Theorems \ref{ix} and \ref{or},  we obtain the following.

\bthm\label{wa}
Let $\xi$ be a solution of the $j$-th $n\times n$-KdV flow \eqref{in}, $Y=$ $\sum_{i=0}^{n-1} y_i \L^i$ a solution of \eqref{os}, and $\ti \xi$ the solution of \eqref{in} constructed from $\xi$ and $Y$ as in Theorem \ref{or}. Let $u(\cdot, t)= \Psi(\xi(\cdot, t))$, $\ti u=\Psi(\ti\xi(\cdot, t))$, $\D(\cdot, t)= \G(\xi(\cdot, t))$, and $\ti \D(\cdot, t)= \G(\ti\xi(\cdot, t))$, where $\Psi$ and $\G$ are defined in Definition \ref{sm}. Set $$f= \ti \D (J+Y) \D^{-1}.$$ Then we have the following:
\ben
\item[(a)] Both $u$ and $\ti u$ are solutions of the $j$-th $\an1$-KdV flow \eqref{dga}.
\item[(b)] If $F$ is a frame of $\xi$, then $E= F\D^{-1}$ and $\ti E= Ef^{-1}$ are frame of solutions $u$ and $\ti u$ of \eqref{dga} respectively.
\item[(c)] $f$ is of the form $J+ y_0\I_n+ N$, where $N$ is an $\calN_n^+$-valued map.
\een  
\ethm

\begin{proof}
From Theorem \ref{ix}, both $u$ and $\ti u$ are solutions of \eqref{dga}, $E=F\D^{-1}$ and $\ti E=\ti F \ti \D^{-1}$ are frames of $u$ and $\ti u$ respectively. It follows from a direct computation that 
$$\ti F\ti\D^{-1}=F(J+Y)^{-1}\ti \D^{-1}= E\D (J+Y)^{-1}\ti \D^{-1}= Ef^{-1}.$$
This proves (a) and (b).

Let $\calG_i=\oplus_{j=1}^{n-i} \C e_{j, i+j}$. Write $\D= \sum_{i=0}^{n-1} \D_i$ and $\ti \D= \sum_{i=0}^{n-1} \ti \D_i$ with $\D_i$, $\ti \D_i\in \calG_i$. By Theorem \ref{ix}, $\D_1= \ti \D_1=0$.  So $f= J+y_0\I_n+ N$ for some $N(x,t)\in \calN_n^+$.  
\end{proof}

Next we prove that entries of $N$ in Theorem \ref{wa} are differential polynomials of $u$ and $h$. First we need a Lemma, which can be proved by a direct computation. 

\blem\label{ni}  Let $u=\sum_{i=1}^{n-1}u_ie_{in}$, $E \in C^{\infty}(\R, GL(n, \C))$ and $\phi$ the first column of $E$. Then $E_x= E(J+u)$ if and only if $E=(\phi, \phi_x, \ldots, \phi_x^{(n-1)})$ and $\phi_x^{(n)}= \sum_{i=1}^{n-1} u_i \phi_x^{(i-1)}+\l\phi$, where $J=e_{1n}\l+b$ is defined by \eqref{kt}. 
\elem

We say a differential polynomial $\eta$ {\it has order $k$ in $h$\/} if $\eta$ is a polynomial in $h, h_x, \ldots, h_x^{(k)}$.  

\bthm\label{mz} Let $E$ and $\ti E$ be frames of solutions $u=\sum_{i=1}^{n-1}u_ie_{in}$ and $\ti u=\sum_{i=1}^{n-1}\ti u_ie_{in}$ of \eqref{dga} respectively, and $\phi$, $\ti \phi$ the first column of $E$ and $\ti E$ respectively. Suppose there exists $h(x, t)$ such that $\phi=(\p_x+h)\ti \phi$. Then $E(x, \l)=\ti E(x, \l)f$, where $f$ is of the form $J+hI_n+N$, for some $N(x, t) \in \cn_n^+$. Moreover,
\ben
\item[(i)] there are differential polynomial $s_i(u,h)$ of order $(n-i)$ in $h$ such that
\beq\label{nm1}
\ti u_i= u_i + s_i(u,h), \quad 1\leq i\leq n-1,
\eeq
\item[(ii)]
the $ij$-th entry of $N$ is
\beq\label{jh1}
\bca N_{ij}= C_{j-1, i-1} h_x^{(j-i)}, & 1 \leq i < j< n,\\
N_{in}=u_i + s_i(u,h) + C_{n-1, i-1} h_x^{(n-i)}, & 1\leq i\leq n-1,\eca
\eeq
where $C_{j,i}=\frac{j!}{i!(j-i)!}$.  (We will use $f_{u,h}$ to denote such $f$),
\item[(iii)] $h$ satisfies 
\beq\label{btj}
\bca h_x^{(n)} = r_n(u,h),\\
h_t= \eta_{n,j}(u,h),\eca
\eeq
for some differential polynomials $r_n(u, h)$ of order $(n-1)$ in $h$ and $\eta_{n,j}(u,h)$ of order $j$ in $h$.
\een
\ethm

\begin{proof} From $E^{-1}E_x=J+u$, $\ti E^{-1}\ti E_x=J+\ti u$, and $E= \ti Ef$, we have 
\beq\label{cm1}
f_x =  f(J+u)- (J+\ti u) f.
\eeq
Compare the $nn$-th entry of \eqref{cm1} to get 
$u_{n-1}= h_x +\ti u_{n-1}+(n-1)h_x$, i.e., 
\beq\label{kw}
\ti u_{n-1} = u_{n-1}-nh_x.
\eeq Compare the $(n-1, n)$-entry of \eqref{cm1} to get 
\beq\label{kwa}
\ti{u}_{n-2}=u_{n-2}-(u_{n-1})_x-\frac{(n-3)n}{2}h_{xx}+nhh_x.
\eeq
Use induction and compare the $(n-i, n)$-th entry of \eqref{cm1} for $1\leq i\leq n-1$  to see that  $\ti u_{n-i}= u_{n-i} + s_{n-i}(u,h)$ for some differential polynomial $s_{n-i}(u,h)$ of order $i$ in $h$.  This proves (i). 

Let $f=\ti E^{-1} E$. By assumption, the first column of $f$ is $(h,1,0, \ldots,0)^t$.  
Lemma \ref{ni} implies that $\ti \phi_x^{(n)}= \l \ti\phi + \sum_{i=1}^{n-1} \ti u_i \ti\phi_x^{(i-1)}$. Use $\phi= \ti\phi_x+ h \ti \phi$ to compute $\phi_x^{(i)}$ and use \eqref{nm1} to get \eqref{jh1}. This proves (ii).

Compare the $1n$-th entry of the constant term of \eqref{cm1} (as a polynomial in $\l$) to get $h_x^{(n)}= r_n(u,h)$ for some differential polynomial $r_n(u,h)$ of order $(n-1)$ in $h$. Since $E$ and $\ti E$ are frames of $u$ and $\ti u$, we have $E^{-1}E_t=Z_j(u, \l)$ and $\ti E^{-1}\ti E_t=Z_j(\ti u, \l)$, where $Z_j$ is as in \eqref{pp}. But $E=\ti E f$ implies that 
\beq\label{cm2}
f_t= fZ_j(u,\l) - Z_j(\ti u,\l) f. 
\eeq
Compare the $11$-th entry of the constant term of \eqref{cm2} to see that $h_t=\eta_{n,j}(u,h)$ for some differential polynomial $\eta_{n,j}(u,h)$ of order $j$.  This proves (iii).
\end{proof}

The proof of Theorem \ref{mz} gives the following Proposition in $x$-variable:
\bprop\label{mzb} Let $u=\sum_{i=1}^{n-1}u_ie_{in}$, $\ti u=\sum_{i=1}^{n-1}\ti u_ie_{in} \in C^{\infty}(\R, V_n)$, $E, \ti E \in C^{\infty}(\R, GL(n, \C))$ satisfying $E^{-1}E_x=J+u$ and $\ti E^{-1} \ti E_x=J+\ti u$. Suppose $E=\ti E f$ for some $f$ of the form $J+h\I_n+N$, where $h\in C^\infty(\R, \C)$ and $N \in C^{\infty}(\R, \cn_n^+)$.  Then $f=f_{u, h}$, $\ti u_i=u_i+s_i(u, h)$ for $1 \leq i \leq n-1$, and $h_x^{(n)}=r_n(u, h)$ as in Theorem \ref{mz}.
\eprop
\bprop\label{mf} If $g(x)\in GL(n,\C)$ and $\tr(g^{-1}g_x)=0$, then $\det(g(x))$ is a constant.
\eprop
\begin{proof} Since $\frac{\rd}{\rd x} \ln(\det(g))= \tr(g^{-1}g_x)$, $\det(g)$ is constant.
\end{proof}

\bprop\label{ml} Let $u, \ti u$ and $E, \ti E$ be as in Proposition \ref{mzb}. Then 
 there exists a differential polynomial $\xi_n(u,h)$ of order $(n-2)$ in $h$ such that 
\begin{align} &\det(f_{u,h}(x,\l))=(-1)^{n-1}(\l+ h_x^{(n-1)}- \xi_n(u,h)), \label{pv} \\
& h_x^{(n)}-r_n(u,h)= (h_x^{(n-1)}-\xi_n(u,h))_x, \label{pw}
\end{align}
where $f_{u,h}$ and $r_n(u,h)$ are as in Theorem \ref{mz}. 
\eprop

\begin{proof} 
Use \eqref{jh1} and a direct computation to get \eqref{pv}. Let $f_{u, h}=f$. The proof of Theorem \ref{mz} implies that 
\beq\label{ky}
f_x- f(J+u) + (J+\ti u) f= (h_x^{(n)}- r_n(u, h))e_{1n}.
\eeq
Let $w= h_x^{(n)}- r_n(u,h)$.  By \eqref{ky}, we get
 $$(\ln(\det f))_x = \tr(f^{-1}f_x) =\tr((J+u)-f^{-1}(J+\ti{u})f+f^{-1}we_{1n}).$$
 Since $\tr(J+u)=\tr(J+\ti u)=0$, 
 $$(\ln(\det f))_x= \tr(f^{-1} we_{1n})=wf^{n1},$$
 where $f^{n1}$ is the $(n1)$-th entry of the inverse of $f$.  By definition of $f$, we see that $f^{n1}$ is equal to $(-1)^{n-1}(\det f)^{-1}$. This proves \eqref{pw}.
\end{proof} 

\bthm\label{ot}  Suppose $E, \ti E$ are frames of solutions $u, \ti u$ of the $j$-th $\an1$-KdV flow \eqref{dga} respectively, and $\ti E= E f_{u,h}^{-1}$ for some smooth function $h$.  Then there exists a constant $k\in \C$ such that $\det(f_{u, h}(x, t, k))=0$ and $h$ satisfies
\beq\label{btn}
{\rm (BT)}_{u,k}\quad \bca h^{(n-1)}_x= \xi_n(u,h)-k,\\ h_t= \eta_{n,j}(u,h),\eca  
\eeq
where $\xi_n(u,h)$ is as in Proposition \ref{ml} and $\eta_{n,j}(u,h)$ is as in Theorem \ref{mz}.  
\ethm

\begin{proof} Recall that $\tr(J+u)=\tr(J+\ti u)= \tr(Z_j(u,\l))= \tr(Z_j(\ti u,\l))=0$. By Proposition \ref{mf},  determinants of both $E(x,t,\l)$ and $\ti E(x,t,\l)$ are independent of $x$ and $t$.  Hence $\det(f_{u,h})$ is independent of $x, t$. But $\det(f_{u,h})$ is a degree one polynomial in $\l$. So there exists a constant $k\in\C$ such that $\det(f_{u,h}(x,t,\l))= (-1)^{n-1}(\l-k)$. By \eqref{pv},  we have $\det(f_{u,h})=(-1)^{n-1} (\l + h_x^{(n-1)} -\xi_n(u,h))$. Hence
$$h_x^{(n-1)}-\xi_n(u,h)=-k.$$ 
It follows from Theorem \ref{mz} that $h$ satisfies the second equation of \eqref{btn}. 
\end{proof}

The following Theorem gives a method to construct solutions of (BT)$_{u, k}$ from frames of $u$.
\bthm \label{wb} Let $E$ be a frame of a solution $u=\sum_{i=1}^{n-1}u_ie_{in}$ of the $j$-th $\an 1$-KdV flow \eqref{dga}, $k \in \C$, $\bc_0 \in \C^n \bh 0$ constants, and $v(x, t)=(v_1, \cdots, v_n)^{t}(x, t):=E(x, t, k)^{-1}(\bc_0)$. If $v_n \neq 0$, then
\ben
\item[(i)]  $h=-\frac{v_{n-1}}{v_n}$ is a solution of {\rm (BT)$_{u, k}$},
\item[(ii)] $\ti E=Ef_{u, h}^{-1}$ is a frame of a new solution $\ti u= \sum_{i=1}^{n-1} \ti u_i e_{in}$ of \eqref{dga}, where $\ti u_i$'s are given by \eqref{nm1},
\item[(iii)] $f_{u,h}(x,t,k)(v(x,t))=0$ for all $x, t$. 
\een 
\ethm

\begin{proof} By Theorem \ref{ix}, there exist a unique solution $\xi$ of the $j$-th $n\times n$ KdV flow \eqref{in} such that $\Psi(\xi(\cdot, t))= u(\cdot, t)$ and $\D(\cdot,t)= \G(\xi(\cdot, t)) \in N_n^+$, where $\Psi$ and $\G$ are as in Definition \ref{sm}.  Then $F= E\D$ is a frame for $\xi$.

Let $\zeta=(\zeta_1, \cdots, \zeta_n)^t=\D^{-1}v$, so $\zeta=F(x, t, k)^{-1}(\bc_0)$.  Let $Y(\zeta)$ be as in Theorem \ref{or} (iii). Then $\ti F=F(J+Y)^{-1}$ is a frame of a new solution $\ti \xi $ of \eqref{in}, $y_0=-\frac{\zeta_{n-1}}{\zeta_n}$, and $(ke_{1n}+b+Y)\zeta=0$. Hence $\det(J+Y)(x, t, k)=0$. By Theorem \ref{wa}, $\ti u=\Psi(\ti \xi)$ is a solution of $\eqref{dga}$ and $\ti E=E f^{-1}$ is a frame of $\ti u$, where 
$f=\ti \D(J+Y)\D^{-1}$. By Theorem \ref{mz}, $f=f_{u, y_0}$. Therefore, $\det(f_{u, y_0})(x, t, k)=0$. By Theorem \eqref{ot}, $y_0$ is a solution of \eqref{btn}. Write $\D= \I_n+\sum_{i=1}^{n-1}\D_{i}$ and $\ti \D= \I_n+ \sum_{i=1}^{n-1}\ti \D_i$ with $\D_i\in \calG_i=\oplus_{j=1}^{n-i} \C e_{j, j+i}$.  By Theorem \ref{ix} (b), we have $\D_1=\ti \D_1=0$. It follows from a  direct computation that $y_0=-\frac{\zeta_{n-1}}{\zeta_n}=-\frac{v_{n-1}}{v_n}=h$.  So we have proved (i) and (ii).

From $(ke_{1n}+b+Y)(\zeta)=0$, $\zeta=\D^{-1}v$ and $f_{u, h}=\ti \D (J+Y) \D^{-1}$, we have $f_{u, h}(x, t, k)v(x, t, k)=0$.
\end{proof}

\bthm\label{nb} Let $u$ be a solution of \eqref{dga}, and $k\in \C$ a constant. Then
\ben
\item[(i)] all solutions of {\rm (BT)$_{u,k}$} are obtained by the method given in Theorem \ref{wb}.
\item[(ii)] {\rm (BT)$_{u,k}$}, i.e.,  \eqref{btn} with parameter $k$, is solvable,
\een
\ethm

\begin{proof} Let $E$ be the frame of a solution $u$ of \eqref{dga} such that $E(0,0,\l)=\I_n$ for all $\l\in \C$. Let $\bc=(c_1, \cdots, c_{n-1}, 1)$ and $v=(v_1, \ldots, v_n):= E^{-1}(\cdot, \cdot, k)(\bc)$. Then $v_n \neq 0$ in an open neighborhood of $(0, 0)$ in $\R^2$. It also follows from $E^{-1}E_x=J+u$ that 
\beq\label{sv}
v_x=-(e_{1n}k+b+u)v.
\eeq

Let $h=-v_{n-1}/v_n$, and $\Phi: \C^{n-1} \rightarrow \C^{n-1}$ the map defined by 
$$\Phi(\bc)=(h(0, 0), h_x(0, 0), \cdots, h_x^{(n-2)}(0, 0)).$$
We claim that $\Phi$ is onto. Let $\phi=-\frac{1}{v_n} v$, then $h=\phi_{n-1}$. Eq.\eqref{sv} implies that 
$$
\bca
\phi_x=-(ke_{1n}+b+u+hI_n)\phi, \\
\phi(0, 0)=-\bc.
\eca
$$
Compare the entries of both sides, we have 
\begin{align*}
& (\phi_1)_x=-\phi_1\phi_{n-1}+k+u_1, \\
& (\phi_i)_x=-\phi_{i-1}-\phi_i\phi_{n-1}+u_i, \quad 2 \leq i \leq n-1.
\end{align*}
This shows that given any $(c_1, \ldots, c_{n-1})$ there exists a solution $h$ of \eqref{btn} satisfying 
$$h(0, 0)=-c_{n-1}, \quad h_x(0, 0)=c_{n-2}+u_{n-1}(0, 0)-c_{n-1}^2, \quad \cdots$$
This proves (i). Since we have constructed solutions with all initial data, system \eqref{btn} with parameter $k$ is solvable. 
\end{proof}

\bcor\label{nba} If $u=\sum_{i=1}^{n-1}u_ie_{in}$ is a solution of \eqref{dga} and $h$ is a solution of {\rm (BT)$_{u, k}$} \eqref{btn}, then $\ti u=\sum_{i=1}^{n-1}\ti u_i e_{in}$ defined by $\ti u_i=u_i+s_i(u, h)$ as in \eqref{nm1} is a solution of \eqref{dga}. Moreover, if $E$ is a frame of $u$, then $E f_{u, h}^{-1}$ is a frame of $\ti u$ for $\l \neq k$ and $\det(f_{u, h}(x, t, \l))=(-1)^{n-1}(\l -k)$.
\ecor

\brem Given a solution $u$ of \eqref{dga}, we can construct new solutions of \eqref{dga} either by solving the compatible non-linear system $({\rm BT})_{u,k}$, i.e., \eqref{btn}, for $h$ or solve the frame $E(x, t, k)$ by solving a linear system. When $n=2$ this is the classical relation between two component linear systems and systems of Riccati equations. 
\erem

It follows from \eqref{pw} that if $h$ is a solution of \eqref{btj} then $h_x^{(n-1)}-\xi_n(u, h)=k(t)$ for some function of $t$. We need the following lemma to prove that $k(t)$ is a constant. 

\blem\label{rh}  Let $K:\oplus_{i=0}^{n-1}\C\L^i\to \C^n$ be the map defined by 
$$K({\bc})=(y_0(0, 0), (y_0)_x(0, 0), \cdots, (y_0)_x^{(n-1)}(0, 0))^t,$$ 
 where  $Y=\sum_{i=0}^{n-1}y_i\L ^i$ is the solution of \eqref{os} with initial data $Y(0,0)={\bc}$. 
Then $K$ is onto. 
\elem

\begin{proof} Let $F$ be a frame of $\xi=\sum_{i=1}^{n-1}\xi_i \L^i$ of the $j$-th $n \times n$ KdV flow \eqref{in}. By Theorem \ref{or}, $\ti F=F(J+Y)^{-1}$ is a frame for a new solution $\ti \xi=\sum_{i=1}^{n-1}\ti \xi_i \L^i$. Hence we have  
$$ (J+Y)_x=(J+Y)(J+\xi)-(J+\ti \xi)(J+Y).$$
The coefficients of $\l$ of the both sides are zero. So the above equality is equivalent to 
\beq\label{rv}(b+Y)_x=(b+Y)(b+\xi)-(b+\ti \xi)(b+Y).
\eeq
Compare the $(n, n)$-th entry of both sides to get
$$y_1=\a (y_0)_x+\xi_1, \quad \text{where} \ \a = e^{\frac{2\pi i}{n}}.$$
Let $\cg_i=\oplus_{j=1}^{n-i}\C e_{i, j+i}$. Compare the $\cg_i$-components of both sides of \eqref{rv} for $1 \leq i \leq n-2$, and use the fact that 
$$\L^kb-b\L^k=(\sum_{i=1}^{k-1}\a^i)a\L^{k-1}, \quad a =\diag(1, \a, \cdots, \a^{n-1}), \quad \a =e^{\frac{2 \pi i}{n}}.$$
We obtain 
\begin{align*}
(y_i)_x\L^i = & (y_{i+1}-\xi_{i+1})\left(\sum_{j=1}^{i}\a^j\right)a\L^i+(\xi_{i+1}-\ti{\xi}_{i+1})\L^{i+1}b \\
& + \left( \sum_{k=0}^{i}y_k(\xi_{i-k}-\ti \xi_{i-k})\right)\L^i, \quad (\xi_0=\ti \xi_0=0).
\end{align*}
Compare the $(n-i, n)$-th entry of both sides and by induction to get 
$$y_{i+1}=\frac{\a^{i+1}}{\sum_{j=1}^{i}\a^j}(y_{i})_x+p_{i}(y_0, \cdots, y_{i-1}, \xi), \quad 1 \leq j \leq n-2,$$
where $p_i$'s are polynomials in $y_0, \cdots, y_{i-1}$ and $\xi$. Therefore, given $(y_0)_x^{(i)}(0, 0)$ for $0 \leq i \leq n-1$, we can write down $y_i(0, 0)$, $0 \leq i \leq n-1$ uniquely. So K is onto. 
\end{proof}

\bthm\label{ow} Let $u$ be a solution of the $j$-th $\an 1$-KdV flow \eqref{dga}, and $h$ a solution of \eqref{btj}. Let $\ti u_i= u_i+s_i(u, h)$ as in \eqref{nm1}, and  $\ti u= \sum_{i=1}^{n-1}\ti u_ie_{in}$. Then $\ti u$ is a solution of \eqref{dga} and $\ti E=E f_{u, h}^{-1}$ is a frame of $\ti u$. 
\ethm

\begin{proof} 
   By Theorem \ref{ix}, there exist a unique solution $\xi$ of the $j$-th $n\times n$ KdV flow \eqref{in} such that $\Psi(\xi(\cdot, t))= u(\cdot, t)$ and $\D(\cdot,t)=\G(\xi(\cdot, t)) \in N_n^+$ and $F= E\D$ is a frame for $\xi$.

By Lemma \ref{rh}, there exists $\bc\in \oplus_{i=0}^{n-1} \C \L^i$ such that 
$$K(\bc)=(h(0,0), h_x(0,0), \ldots, h_x^{(n-1)}(0,0))^t.$$
Let $Y=\sum_{i=0}^{n-1} y_i \L^i$ be the solution of \eqref{os} with initial data 
$$\sum_{i=0}^{n-1} y_i(0,0)\L^i={\bc}.$$  
By Theorem \ref{or}, $\hat F= F(J+Y)^{-1}$ is a frame of the new solution $\hat \xi$ of \eqref{in}.  Let $\hat u(\cdot, t) =\Psi(\hat \xi(\cdot, t))$, and $\hat\D(\cdot, t)= \G(\hat \xi(\cdot, t))$. By Theorem \ref{ix} again, $\hat u$ is a solution of \eqref{dga} and $\hat E= \hat F\hat \D^{-1}$ is a frame of $\hat u$.  Since $F= E\hat \D$, we see that
$$\hat E= E\D (J+Y)^{-1} \hat \D^{-1} = Eg^{-1},$$
where $g= \hat \D (J+Y) \D^{-1}$. By Theorem \ref{wa}, $g$ is of the form $J+ y_0\I_n + N$ for some $\calN_n^+$-valued map. From Theorem \ref{mz}, $g=f_{u, y_0}$. This shows that $\hat E, E$ are frames of solutions $u, \hat u$ of \eqref{dga} and $\hat E= Ef_{u, y_0}^{-1}$. By Theorem \ref{mz}, $y_0$ satisfies \eqref{btj}.  By assumption, $h$ is  a solution of \eqref{btj}.  We have chosen the initial data of $Y$ for system \eqref{os} so that $(y_0)_x^{(i)}(0,0)= h_x^{(i)}(0,0)$ for $0\leq i\leq n-1$.  So $h$ and $y_0$ are solutions of \eqref{btj} with the same initial values at $(0,0)$.  By Frobenius Theorem, there is only one solution with the same initial data. Hence $y_0=h$.  This proves $\hat E= Ef_{u,h}^{-1}$ is a frame of the new solution $\hat u=\sum_{i=1}^{n-1} \hat u_i e_{in}$ and $\hat u_i=u_i+ s_i(u,h)$.    
\end{proof}

\bcor\label{spa}
Let $E$ be a frame of a solution $u$ of \eqref{dga}. Then $\ti E= Ef_{u,h}^{-1}$ is a frame of some solution of \eqref{dga} if and only if $h$ is a solution of \eqref{btj}. 
\ecor

\bcor\label{spb} 
 If $u$ is a solution of \eqref{dga} and $h$ is a solution of \eqref{btj}, then there exists a constant $k\in \C$ such that $h$ is a solution of $( {\rm BT})_{u,k}$ \eqref{btn}.
\ecor

\bcor\label{sp}  If $u$ is a solution of the $j$-th $\an1$-KdV flow \eqref{dga}, then
\ben
\item[(a)] system \eqref{btj} is solvable,
\item[(b)] the space of solutions of system  \eqref{btj} is the union of the spaces of solutions of $( {\rm BT})_{u,k}$, i.e., \eqref{btn}, for all $k\in \C$.
\een
\ecor

\begin{proof}  From the proof of Theorem \ref{ow}, given any constant $(d_0, \cdots, d_{n-1})^t \in \C^{n}$, there is a solution $h$ of \eqref{btj} such that $h_x^{(i)}(0, 0)=d_i$ for $0 \leq i \leq n-1$. This proves (a). It follows from Theorem \ref{ot} and \ref{ow} that, if $h$ is a solution of \eqref{btj}, then there exists a constant $k \in \C$ such that $h_x^{(n-1)}=\xi_n(u, h)-k$. Hence $h$ is a solution of \eqref{btn} with parameter $k$.
\end{proof} 

\bdefn Let $u$ be a solution of \eqref{dga}, and $h$ a solution of (BT)$_{u,k}$, i.e.,  \eqref{btn}. We use $h \ast u$ to denote the solution $\ti u$ constructed from $u$ and $h$ as in Corollary \ref{nba} and call $u \mapsto h \ast u$ {\it a B\"{a}cklund transformation with parameter $k$\/}.
\edefn

\beg We use the algorithm given in section \ref{km} to compute
the second $A_2^{(1)}$-KdV flow and obtain
\beq\label{jt}
\begin{cases}
(u_1)_t = (u_1)_{xx}-\frac{2}{3}(u_2)_{xxx}+\frac{2}{3}u_2(u_2)_x,\\
(u_2)_t = -(u_2)_{xx}+2(u_1)_x.
\end{cases}
\eeq
Note that solutions of \eqref{jt} give rise to solutions of the {\it Boussinesq equation\/}. In fact, take the derivative with respect to $t$ on both sides of the second equation and use the  first equation to see that $u_2$ satisfies the Boussinesq equation:
\beq\label{bsq}
(u_2)_{tt}=-\frac{1}{3}(u_2)_x^{(4)}+\frac{4}{3}(u_2)_x^2+\frac{4}{3}u_2(u_2)_{xx}.
\eeq
Note that the right hand side is equal to $-\frac{1}{3} (u_2)_{xxx} + \frac{4}{3} (u_2(u_2)_x)_x$. 

We use the proof of Theorem \ref{mz} to compute differential polynomials $s_i(u, h)$, $r_n(u, h)$ and $f_{u, h}$. The  system \eqref{btn} for a solution $u=u_1 e_{13}+ u_2e_{23}$ of \eqref{jt} is
\beq\label{btn3}
\bca 
h_{xx}= -u_1+(u_2)_x + hu_2 -3hh_x - h^3-k,\\
h_t= \frac{2}{3} (u_2)_x - h_{xx} - 2hh_x,
\eca
\eeq
and system \eqref{btj} is 
\beq\label{bt3}
\bca 
h_x^{(3)}= ( -u_1+(u_2)_x + hu_2 -3hh_x - h^3)_x, \\
h_t= \frac{2}{3} (u_2)_x - h_{xx} - 2hh_x.
\eca
\eeq
The new solution  $h\ast u=\ti u_1 e_{13}+ \ti u_2 e_{23}$ of \eqref{jt} is given by 
\beq\label{ct}
\bca \ti u_1= u_1- (u_2)_x + 3hh_x,\\ \ti u_2= u_2-3h_x.\eca
\eeq
Moreover, if $E$ is a frame for $u$, then $Ef_{u,h}^{-1}$ is a frame for $h\ast u$, where
$$
f_{u,h}(x,t,\l)= e_{13}\l + \bpm h& h_x& u_1-(u_2)_x+ h_{xx} + 3hh_x\\ 1 & h & u_2- h_x\\ 0& 1&h\epm.
$$
\eeg

If $u$ is a solution of \eqref{dga} and $h$ is a solution of (BT)$_{u,k}$, then by \eqref{pv}, $\det(f_{u,h}(x,t,k))=0$.  So $Ef_{u,h}^{-1}$ is not holomorphic at $\l=k$.  However, we can multiply $Ef_{u,h}^{-1}$ on the left by some $C(\l)$ independent of $x, t$ to get a new frame for $h\ast u$ that is holomorphic at $\l=k$.  

\bthm\label{pb} Suppose $E(x,t,\l)$ is a frame of a solution $u$ of \eqref{dga} that is holomorphic for $\l$ in an open subset $\calO$ of $\C$. Let $k\in \co$ be a constant, and $h$ the solution of \eqref{btn} constructed from $E(x, t, k)$ and $\bc=(c_1, \cdots, c_n)^t \in \C^n \bh 0$ as in Theorem \ref{wb}.  Then
\ben
\item[(i)] $\ti E= E f_{u,h}^{-1}$ is a frame of the new solution $h\ast u$ and $\ti E$ is holomorphic for all $\l\in \calO\bh \{k\}$.
\item[(iv)] Let  $C(\l)= e_{1n}(\l-k)+ b+ \sum_{i=1}^{n-1} c_i e_{i+1, n}$. Then 
$$\hat E(x,t,\l)= C(\l) E(x,t,\l)f_{u,h}(x,t,\l)^{-1}$$ 
is a frame for $h\ast u$ that is holomorphic for all $\l\in \calO$.  
\een
\ethm

\begin{proof}  By Corollary \ref{sp}, $h$ is a solution of \eqref{btj}.  It follows from Theorem \ref{ow} that $h\ast u$ is a solution of \eqref{dga} and $Ef_{u,h}^{-1}$ is a frame of $h\ast u$.  By definition, $f_{u,h}(x,t,\l)$ is holomorphic for all $\l\in \calO$.  
Since $\det(f_{u,h}(x,t,\l)=(-1)^{n-1}(\l-k)$,  if $\l\not=k$ then $Ef_{u,h}^{-1}$ is holomorphic at $\l$.  

Let $f=f_{u,h}$, and $f^\sharp$ the matrix whose $ij$-th entry is $(-1)^{i+j} \det(M_{ji})$, where $M_{ij}$ is the $(n-1)\times (n-1)$ matrix obtained from $f$ by crossing out the $i$-th row and $j$-th column. Then $ff^\sharp =  (-1)^{n-1}(\l-k) \I_n$.  So we have 
\beq\label{pa1}
\Im(f^\sharp(x,t, k))\subset \Ker(f(x,t,k))= \C E^{-1}(x,t,k)({\bf c}).
\eeq 
Let 
\begin{align*}
F(x,t,\l)&= C(\l)E(x,t,\l) f_{u,h}^{-1}(x,t,\l)\\
& = (-1)^{n-1} \frac{1}{\l-k} C(\l) E(x,t,\l) f^\sharp (x,t,\l).
\end{align*}
Then $F(x,t,\l)$ is holomorphic for $\l\in \calO\bh \{k\}$ and has a possible simple pole at $\l=k$.  We claim that the residue of $F(x,t,\l)$ at $\l=k$ is zero.  
The residue of $F(x,t,\l)$ at $\l=k$ 
is equal to $(-1)^{n-1} C(k) E(x,t,k) f^\sharp(x,t,k)$.  By \eqref{pa1}, it is equal to 
$\phi(x,t) C(k) (\bc)$ for some function $\phi$. A direct computation implies that $C(k)(\bc)=0$. This proves the claim and  $F$ is holomorphic at $\l=k$.  
\end{proof}

The above Theorem will be used to construct B\"acklund transformations for the n-d central affine curve flow in \cite{TWb}.

\bs
\section{Permutability and Scaling Transform}\label{btp}
In this section, we 
\ben
\item[(i)] give a Permutability formula for B\"acklund transformations, 
\item[(ii)] prove that the conjugation of a B\"acklund transformation with parameter $k=1$ by the $r$-scaling transform gives a BT with parameter $k=r^{-n}$
\een
for the $\an1$-KdV hierarchy.

\bthm {\rm [Permutability for BT]}\  \label{pa}

\ni Let $u$ be a solution of \eqref{dga}, $k_1, k_2\in \C$ constants,  $h_i$ solutions of {\rm (BT)$_{u,k_i}$} \eqref{btn}, and $h_i\ast u$ the solution of \eqref{dga} construct from $u$ and $h_i$ for $i=1,2$. Suppose $h_1 \neq h_2$. 
Set
\beq \label{if}
\bca
\ti{h}_1=h_1+\frac{(h_1-h_2)_x}{h_1-h_2}, \\
\ti{h}_2=h_2+\frac{(h_1-h_2)_x}{h_1-h_2}.
\eca
\eeq
Then
\ben
\item[(i)] $\ti h_1$ is a solution of {\rm (BT)$_{h_2\ast u, k_1}$} and $\ti h_2$ is a solution of {\rm (BT)$_{h_1\ast u, k_2}$},
\item[(ii)] $\ti h_1\ast (h_2 \ast u)= \ti h_2\ast(h_1\ast u)$. 
\een
\ethm

\begin{proof} Let $E(x,t,\l)$ be a frame of $u$. By Theorem  \ref{nb}, there exist constant vectors  $v_1^0, v_2^0\in \R^n$ that give $h_1, h_2$, i.e., 
\begin{align*}
&v_1=(v_{1,1}. \cdots, v_{1, n})^t=E(\cdot, \cdot, k_1)^{-1}v_1^0, \\
&v_2=(v_{2,1}. \cdots, v_{2, n})^t=E(\cdot, \cdot, k_2)^{-1}v_2^0, \\
& h_i=-\frac{v_{i, n-1}}{v_{i, n}}, \quad i=1,2.
\end{align*}
From \eqref{sv}, 
$$
(v_i)_x=-E^{-1}E_x(x,t, k_i)v_i=-(J+u)\mid_{\l=k_i}v_i, \quad i=1, 2.
$$
In particular, we have 
$$
\bca
(v_{i, n-1})_x=-v_{i, n-2}-u_{n-1}v_{i, n}, \\
(v_{i, n})_x=-v_{i, n-1},
\eca
\quad i=1, 2.
$$
Therefore,
\beq \label{ih}
(h_i)_x=\frac{v_{i, n-2}}{v_{i, n}}-h_i^2+u_{n-1}, \quad i=1, 2.
\eeq
By Theorem \ref{wb}, $E_i(x,t,\l)=E(x,t,\l)f_{u, h_i}^{-1}$ is a frame for $h_i\ast u$ for $i=1, 2$ respectively.
Let 
$$
\ti{v}_1=E_2^{-1}(x, t, k_1)v_1^0, \quad \ti{v}_2=E_1(x,t, k_2)^{-1}v_2^0.
$$
Now we compute $$\ti{h}_i=-\frac{\ti{v}_{i, n-1}}{\ti{v}_{i, n}}.$$
From
$$
\ti{v}_1=E_2^{-1}(x, t, k_1)v_1^{0}=f_{u, h_2}(k_1)E(x, t, k_1)^{-1}v_1^0=f_{u, h_2}(k_1)v_1,
$$
we  get 
$$
\bca
\ti{v}_{1, n-1}=v_{1, n-2}+h_2v_{1, n-1}+(u_{n-1}-(h_2)_x)v_{1, n}, \\
\ti{v}_{1, n}=v_{1, n-1}+h_2v_{1, n}.
\eca
$$
Together with \eqref{ih}, we have
$$
\ti{h}_1=-h_2+\frac{v_{1, n-2}v_{2, n}-v_{2, n-2}v_{1, n}}{v_{2, n-1}v_{1, n}-v_{1, n-1}v_{2, n}}=h_1+\frac{(h_1-h_2)_x}{h_1-h_2}.
$$
Similarly, $\ti{h}_2=h_2+\frac{(h_1-h_2)_x}{h_1-h_2}$.

A direct computation implies $f_{u_2, \ti h_1} f_{u, h_2}= f_{u_1, \ti h_2} f_{u, h_1}$. Hence $\ti h_2\ast (h_1\ast u)= \ti h_1\ast(h_2\ast u)$.
\end{proof}

Next we review the {\it scaling transform} of the $\an1$-KdV hierarchy. 

\bprop\label{nv} (\cite{BS93})  Let $u=\sum_{i=1}^{n-1}u_i e_{in}$ be a solution of the $j$-th $\an1$-KdV flow, and $r\in \R\bh \{0\}$. Let $D(r)= \diag(1, r, \ldots, r^{n-1})$, and 
$$(r\cdot u_i )(x,t) = r^{n+1-i} u_i(rx, r^jt), \quad 1\leq i\leq n-1.$$ 
Then 
\ben
\item[(i)] $(r\cdot u)= \sum_{i=1}^{n-1} (r\cdot u_i) e_{in}$ is a solution of the $j$-th $\an1$-KdV flow,
\item[(ii)] if $E(x,t,\l)$ is a frame of $u$ then 
$$\ti E(x,t,\l):= D(r)^{-1} E(rx, r^j t, r^{-n} \l) D(r)$$ is a frame of $r\cdot u$. 
\een
\eprop

So the multiplicative group $\R\bh \{0\}$ acts on the space of solutions of the $j$-th $\an1$-KdV flow. The following Theorem proves that the conjugation of BT with parameter $k=1$ by a scaling transform gives BT with arbitrary non-zero real parameter.

\bthm\label{nw}  Let  $u$ be a solution of the $j$-th $\an1$-KdV flow, $r \in \R \bh \{0\}$, and $h$ a solution of {\rm (BT)$_{r\cdot u,1}$}. Then 
\ben
\item[(i)] $\hat h(x,t)= r^{-1}h(r^{-1}x, r^{-j}t)$ is a solution of {\rm (BT)$_{u, r^{-n}}$},
\item[(ii)] $r^{-1}\cdot (h \ast (r\cdot u))= \hat h\ast u$. 
 \een
\ethm

\begin{proof} Let $E$ be the frame of $u$ with $E(0, 0, \l)=I_n$. By Proposition \ref{mf},
\beq\label{nw1} 
\det E(x, t, \l)=1.
\eeq
From Proposition \ref{nv}, 
$$F(x,t,\l)= D(r)^{-1} E(rx, r^jt, r^{-n}\l) D(r)$$
is a frame for $\ti u= r\cdot u$, where $D(r)=\diag (1, r, \cdots, r^{n-1})$. It follows from Theorem \ref{ow} that 
$$F_1(x,t,\l) = F(x,t,\l) f_{r \cdot u, h}^{-1}(x,t,\l)$$
is a frame for $h\ast (r\cdot u)$.  Apply the scaling transform given by $r^{-1}$ to $h\ast \ti u$, we have 
\begin{align}\label{nw2}
F_2(x,t,\l) & = D(r) F_1(r^{-1}x, r^{-j} t, r^n \l) D(r)^{-1} \notag \\
& =E(x,t,\l) D(r) f_{\ti u, h}^{-1}(r^{-1}x, r^{-j}t, r^n\l) D(r)^{-1}.
\end{align}
is a frame for $\hat u = r^{-1}\cdot (h\ast (r\cdot u))$.  A direct computation implies that
\begin{align*}
&D(r) f_{\ti u, h}(r^{-1}x, r^{-j}t, r^n\l) D(r)^{-1} =  \\
&\quad r(e_{1n}\l+ b+ r^{-1} h(r^{-1}x, r^{-j}t)\I_n+r^{-1}D(r) ND(r)^{-1}),
\end{align*}
where $f_{\ti u, h}= e_{1n}\l + b + h\I_n + N$ and $N$ is strictly upper triangular. 

So we have proved $F_2(x, t, \l)=r^{-1}E(x, t, \l)g^{-1}(x, t, \l)$ is a frame of $\hat u=r^{-1} \cdot (h \ast (r \cdot u))$, where
$$
g(x, t, \l)=e_{1n}\l+b+\hat h(x, t)I_n+r^{-1}D(r)N(r^{-1}x, r^{-j}t)D(r)^{-1}.
$$
Note that $r^{-1}D(r)N(r^{-1}x, r^{-j}t)D(r)^{-1} \in \cn_n^+$, and 
\beq\label{nw3}
\hat E(x, t, \l)=r F_2(x, t, \l)=E(x, t, \l)g^{-1}(x, t, \l)
\eeq
is a frame of $\hat u$. So $g=f_{u, \hat h}$.

By Theorem \ref{ot}, we have $\det(f_{\ti u, h})= (-1)^{n-1}(\l-1)$. It follows from \eqref{nw1}, \eqref{nw2} and \eqref{nw3} that 
\begin{align*}
\det (f_{u, \hat h}^{-1}) & =\det (\hat E(x, t, \l))=r^n \det (F_2(x, t, \l)) \\
& =r^n \det (f_{\ti u, h}^{-1}(r^{-1}x, r^{-j}t, r^n\l)) \\
&=(-1)^{n-1}\frac{r^{n}}{r^n\l-1}=\frac{(-1)^{n-1}}{\l -r^{-n}}.
\end{align*}
Hence $\det (f_{u, \hat h})=(-1)^{n-1}(\l-r^{-n})$.
\end{proof}

\bs
\section{Explicit solutions}\label{ns}

In this section we apply Theorem \ref{pb} to $u=0$ to construct explicit solutions of \eqref{dga}.

\bprop \label{rn} If we apply BTs with non-zero parameters to the vacuum solution $u=0$ of the $j$-th $\an1$-KdV flow \eqref{dga} repeatedly, then we obtain infinitely many families of explicit solutions of \eqref{dga}  that are rational functions of exponential functions.  
\eprop

\begin{proof}
Let $\l=z^n$, $\a=e^{\frac{2 \pi i}{n}}$,  
\begin{align*}
& D(z)= \diag(1, z, \ldots, z^{n-1}),\\
& \Xi= (\a^{(i-1)(j-1)}), \\
& A_i(x, t, z)=\exp( \a^{i-1} zx+ (\a^{i-1} z)^j t),\\ 
&(m_1(x,t, z), \ldots, m_n(x,t, z))
= (e^{A_1}, \ldots, e^{A_n})\Xi.
\end{align*}
A direct computation implies that the vacuum frame is
$$E(x,t,z^n)= e^{xJ+tJ^j}=\frac{1}{n}D(z)^{-1} \bpm m_1 & m_2 & \cdots & m_n\\ m_n & m_1 & \cdots & m_{n-1}\\ \vdots &\vdots &\vdots & \vdots\\ m_{2} & m_3 & \cdots & m_1\epm D(z).$$

Given constants $k \neq 0$ and $\bc=(c_1, \cdots,  c_n)^t \in \C^n$, let 
\beq\label{pc}
(v_1, v_2, \cdots, v_{n})^t=E(x, t, k^n)^{-1}(\bc), \quad h=-\frac{v_{n-1}}{v_n}.
\eeq Note that 
$$E(x,t,k^n)^{-1}= E(-x, -t, k^n).$$
Let $\{e_1, \ldots, e_n\}$ be the standard basis of $\C^{n\times 1}$, then 
\begin{align*}
(m_3, \ldots, m_n, m_1, m_2) &= (e^{A_1}, \ldots, e^{A_n})\Xi\pi_1, \\ 
(m_2, m_3, \ldots, m_n, m_1)&= (e^{A_1}, \ldots, e^{A_n})\Xi \pi_2.
\end{align*}
where $\pi_1= (e_3, \ldots, e_n, e_1, e_2)$, $\pi_2= (e_2, e_3, \ldots, e_n, e_1)$ (permutation matrices). Hence  
\begin{align*}
v_{n-1}(x,t)&= k^{-(n-2)} (e^{A_1(-x,-t,k)}, \ldots, e^{A_n(-x,-t, k)})\Xi \pi_1 D(k) \bc,\\
v_n(x,t)&= k^{-(n-1)} (e^{A_1(-x,-t, k)}, \ldots, e^{A_n(-x, -t, k)})\Xi \pi_2 D(k) \bc.
\end{align*}
This implies that $h$ given by \eqref{pc} is a rational function of exponentials. Hence entries of the new solution $h \ast 0$ are rational functions of exponential functions. It follows from Theorem \ref{pb} that, the entries of the frame of  $h\ast 0$, 
$$\ti E(x,t,\l)= C(\l) E(x,t,\l) f_{u,h}(x,t,\l),$$
are given by rational functions of exponentials, where $C(\l)= e_{1n}(\l-k)+ b+ \sum_{i=1}^{n-1} c_i e_{i+1, n}$.  If we apply Theorem \ref{pb} to $h\ast 0$ with parameter $k_2\not=0$, then we get another family of new solutions that are rational of exponential functions. 
\end{proof}
 
 \beg[Explicit solutions of the $j$-th $\an1$-KdV flow]\hfil
 
 We use the same notation as in the proof of Proposition \ref{rn}. 
 First note that $\Xi\pi_2 D(k)$ is invertible. So given any ${\bf a}\in \C^{n\times 1}$, we can find suitable $\bc\in \C^n$ such that $\Xi\pi_2 D(k)\bc = {\bf a}$.  We will write down some explicit solutions for the $j$-th $\an1$-KdV flow \eqref{dga} below.

Choose $\bc$ such that $\Xi \pi_2D(k)\bc= e_1+ e_2$. A simple computation implies that
$\Xi \pi_1 D(k) \bc= e_1+ \a e_2$.   The function $h$ defined by \eqref{pc} is
\beq\label{pg} h=-k \frac{e^{-(kx+k^jt)}+\a e^{-(\a kx+\a^j k^j t)}}{e^{-(kx+k^jt)}+ e^{-(\a kx +\a^j k^j t)}}.\eeq
This is a solution of \eqref{btn} with parameter $k\in \C$.  Let $A, B$ be defined by 
$$\bca A+B= -(kx+ k^j t), \\ A-B= -(\a kx+\a^j k^jt).\eca$$
Then \eqref{pg} becomes $h= -k\frac{e^{A+B}+\a e^{A-B}}{e^{A+B}+ e^{A-B}}$, so we get 
\beq\label{pf} 
h= \frac{k}{2} \left((1-\a) \tanh B - (1+\a)\right), 
\eeq
where $\a=e^{\frac{2\pi i}{n}}$ and $B=\frac{1}{2}((1-\a)k x + (1-\a^j)k^j t)$.
Note that $h$ defined by \eqref{pf} is a solution of \eqref{btn} with parameter $k$. 

\ms
(i) If $k= (1+\bar \a)\mu$ for some $\mu\in \R$, then $h$ given by \eqref{pf} becomes a real function,
\beq\label{pej}
h=-\mu\sin (\frac{2 \pi}{n}) \tan \theta-(1+\cos(\frac{2\pi}{n}))\mu,
\eeq 
where $\theta=\mu\sin (\frac{2 \pi}{n})x+\sum_{i=0}^{j-1}C_{j, i}\mu^j\sin(\frac{2(j-i)\pi}{n})t$. 
Hence 
$$h\ast 0= 3h_x(he_{13} - e_{23})$$  is a real solution of the $j$-th $\an1$-KdV flow. 

In particular, when $j=2$, \eqref{pej} becomes 
\beq\label{pe}
h= -\mu\sin\g \tan(\mu\sin\g (x+ 2(1+\cos\g) \mu t))- (1+\cos\g)\mu,
\eeq
where $\g=\frac{2\pi}{n}$, and $\mu \in \R$. Then $h\ast 0=3h_x(he_{13} - e_{23})$ is a real solution of the second $\an1$-KdV flow. 

\ms
(ii) If $n=2m$ and $j$ is odd, we choose $\bc$ such that $\Xi \pi_2 D(k){\bc}=e_1+e_{m+1}$. Then $\Xi \pi_1 D(k){\bc} =e_1-e_{m+1}$, and $h$ defined by \eqref{pc} is 
$$
h=k\tanh (kx+k^j t).
$$
When $k \in \R \bh 0$, we obtain soliton solutions for the $j$-th $A_{2m-1}^{(1)}$-KdV flow. For example, when $m=1$ and $j=3$, this give rise to the 1-soliton solutions of the KdV equation.
\ms

(iii) If $n=2m$ and $j$ is even, we choose $\bc$ such that $\Xi \pi_2D(k) \bc= e_1+ e_{m+1}$. Then $\Xi \pi_1 D(k) \bc= e_1-e_{m+1}$, and $h$ defined by \eqref{pc} is 
$$h= k \tanh (kx).$$
This gives stationary smooth solutions $h\ast 0$ for the $j$-th $A_{2m-1}^{(1)}$-KdV flow. 
\eeg

\beg [Explicit solutions for the second $A_2^{(1)}$-KdV flow \eqref{jt}]  \hfill

For $n=3$, we have $\a=e^{\frac{2\pi i}{3}}$. 

\ben
\item[(i)]  Let $\frac{k}{2}(1-\a)= -\mu$. Then $h$ defined by \eqref{pf} becomes 
\beq\label{yy}
h=\sqrt{3}\mu\tanh(\sqrt{3}\mu(x-2\mu it))+\mu i.
\eeq
Substitute \eqref{yy} to \eqref{ct} with $u=0$ to see that
$$
\bca
u_1=9\mu^3\sech^2(\sqrt{3}\mu(x-2\mu it))(\sqrt{3}\tanh(\sqrt{3}\mu(x-2\mu it))+i), \\
u_2=-9\mu^2\sech^2(\sqrt{3}\mu(x-2\mu it))
\eca
$$
is a complex solution of \eqref{jt}. Note that $u_2$ is the solution of the Boussinesq equation \eqref{bsq} obtained in \cite{DTT}. 

\item[(ii)] $h$ given by  \eqref{pe} is 
$$h=-\frac{\sqrt{3}}{2} \mu \tan (\frac{\sqrt{3}}{2}\mu(x+\mu t)) -\frac{\mu}{2}.$$
Let $\nu= \frac{\mu}{2}$. Then
$$h_\nu=-\sqrt{3}\nu\tan(\sqrt{3}\nu(x+2\nu t))-\mu$$ is a real solution of \eqref{jt} and
 $u_1, u_2$ defined by
$$\bca u_1= 9\nu^3 \sec^2 (\sqrt{3} \nu(x+ 2\nu t))(1+\sqrt{3}\tan (\sqrt{3} \nu (x+ 2\nu t))),\\
u_2= 9\nu^2 \sec^2 (\sqrt{3} \nu (x+ 2\nu t)),\eca
$$
is the real solution of \eqref{jt}.
\een
\eeg

\ms
\bprop If we apply BT with parameter $k=0$ to the vacuum solution $u=0$ of the $j$-th $\an1$-KdV flow \eqref{dga} repeatedly, then we obtain infinitely many families of explicit rational solutions of \eqref{dga}
\eprop

\begin{proof} Note that $E(x,t,\l)=e^{xJ+ tJ^j}$ is a frame of the solution $u=0$ of \eqref{dga}.  Since entries of $E(x,t,0)=\exp(bx+ b^jt)$ are polynomials in $x$ and $t$, we see that the solution $h$ of \eqref{btn} constructed  in Theorem \ref{wb} with $k=0$ is a rational solution and $\ti E= C(\l) E f_{0,h}^{-1}$ is a frame of the solution $h\ast 0$.  Apply Theorem \ref{pb} with parameter $k=0$ and a constant vector ${\bc}_1$ to $h\ast 0$ to get another solution.  But $\ti E(x,t,0)=C(0)E(x,t,0) f_{0,h}^{-1}(x,t,0)$ is rational in $x, t$.  So the solution $h_1$ of \eqref{btn} with $u=h\ast 0$ is rational.  Hence $h_1\ast(h\ast 0)$ is a rational function.  
\end{proof}
 
\beg \label{rss}{\rm [Rational solutions for the second $A_2^{(1)}$-KdV flow]}\par

The coefficients of the constant term and $\l^{-1}$ of the frame $E(x,t,\l)=e^{xJ+tJ^2}$ of the vacuum solution $u=0$ as a power series in $\l$ are
\begin{align*}
E_0(x,t)&=\exp(bx+ b^2 t)= \bpm 1&0&0\\ x& 1&0\\ \frac{x^2}{2} + t & x& 1\epm,\\
E_1(x, t)&=\bpm xt+\frac{1}{6}x^3 & t+\frac{1}{2}x^2 & x \\ \frac{1}{2}t^2+\frac{1}{2}x^2t+\frac{1}{24}x^4 & xt+\frac{1}{6}x^3 & t+\frac{1}{2}x^2 \\ \frac{1}{2}xt^2+\frac{1}{6}x^3t+\frac{1}{5!}x^5 & \frac{1}{2}t^2+\frac{1}{2}x^2t+\frac{1}{24}x^4 & xt+\frac{1}{6}x^3\epm.
\end{align*}
We apply Theorem \ref{pb} to $u=0$ and $v_{0}=(a_1, a_2, 1)^t$ to get new solutions:
\ben
\item[(i)]
$h=\frac{a_1x-a_2}{1+a_1(\frac{x^2}{2}-t)-a_2x}$
is a solution of \eqref{bt3}.
\item[(ii)] $h\ast 0=u_1 e_{13}+ u_2 e_{23}$ is a rational solution of \eqref{jt}, where 
$$
\bca
u_1=-\frac{3(a_1x-a_2)(\frac{1}{2}a_1^2x^2-a_1a_2x+a_1^2t+a_2^2-a_1)}{(1+a_1(\frac{x^2}{2}-t)-a_2x)^3},\\
u_2=\frac{3(\frac{1}{2}a_1^2x^2-a_1a_2x+a_1^2t+ba_2^2-a_1)}{(1+a_1(\frac{x^2}{2}-t)-a_2x)^2}.
\eca
$$
\een
\eeg

\beg {\rm [Rational solutions for the second $\an1$-KdV flow]} \par 

Recall that  $E(x, t, \l)=e^{xJ+tJ^2}$ is a frame of $u=0$. The coefficients of the constant and $\l$ of $E(x,t,\l)$ as a power series in $\l$ are $E(x,t,0)=\exp(bx+ b^2t)$ and $E_1(x,t)= \frac{\p}{\p\l}|_{\l=0} \exp(xJ + tJ^2)$. A direct computation implies that 
$J^k= (b^t)^{n-k} \l + b^k$, $J^n= \l \I_n$, and $\frac{\p}{\p\l} |_{\l=0}J^k = (b^t)^{n-k}$ for $1\leq k\leq n-1$.  So entries of $E_1(x,t)$ are polynomials in $x, t$.  
In fact, for $n=2m+1$, we get
\begin{align*} E_1(x,t)  &= (I+\sum_{j=1}^{n-1}\frac{x^j}{j !}b^j)(\sum_{j=1}^{m}\frac{t^j}{j !}(b^t)^{n-2j}+\sum_{j=1}^{m}\frac{t^{m+j}}{(m+j) !}b^{2j-1}) \\
 &+(\sum_{j=1}^{n-1}\frac{x^j}{j !}(b^t)^j+\frac{x^n}{n ! }+\sum_{j=1}^{n-1}\frac{x^{n+j}}{(n+j) !}b^j)(I+\sum_{j=1}^m\frac{t^j}{j !}b^{2j}).
\end{align*}
For $n=2m$,
\begin{align*}
 E_1(x,t)  &= (I+\sum_{j=1}^{n-1}\frac{x^j}{j !}b^j)(\sum_{j=1}^{m-1}\frac{t^j}{j !}(b^t)^{n-2j}+\frac{t^m}{m !}+\sum_{j=1}^{m-1}\frac{t^{m+j}}{(m+j) !}b^{2j}) \\
 &+(\sum_{j=1}^{n-1}\frac{x^j}{j !}(b^t)^j+\frac{x^n}{n ! }+\sum_{j=1}^{n-1}\frac{x^{n+j}}{(n+j) !}b^j)(I+\sum_{j=1}^{m-1}\frac{t^j}{j !}b^{2j}).
\end{align*}
Apply Theorem \ref{pb} repeatedly to construct infinitely many families of explicit rational solutions. 
\eeg

\bs 
\section{$Z_n$-action for the KW flows}\label{pia}

In this section we give a natural action of the cyclic group $Z_n$ of $n$ elements on the space of solutions of the $j$-th KW flow and show that  Adler's B\"acklund transformation comes from this $Z_n$-action.

\bthm\label{po} Let $\a= e^{\frac{2\pi i}{n}}$, $a=\diag(1, \a, \ldots, \a^{n-1})$, and $\tau=e_{21} + e_{32} + \cdots + e_{n,n-1}+ e_{1n}$ as in  the KW-hierarchy.  If $v=(v_1, \ldots, v_{n-1})$ is a solution of the $j$-th KW flow \eqref{iq}, then $$\a\cdot v=(\a v_1, \ldots, \a^{(n-1)} v_{n-1})$$ is a solution of \eqref{iq}.  In particular, this defines an $\Z_n$-action on the space of solutions of \eqref{iq}.
\ethm

\begin{proof}  Let $\calL^{KW}= \calL^{KW}_+ \oplus \calL_-^{KW}$ be the splitting that gives the KW-hierarchy in section \ref{km}.  

First note that $a\tau= \a \tau a$.  A direct computation implies that $\Ad(a^{-1})$ leaves $\calL^{KW}_\pm$ invariant, i.e., $a^{-1}(\calL^{KW}_{\pm})a \subset \calL^{KW}_\pm$. So
$$a^{-1}(\p_x+ az + P(v))a= \p_x+ az + P(\a\cdot v).$$
Recall that $\hat Q(v,z)$ is defined by $[\p_x+ a z + P(v), \hat Q(v,z)]=0$ and $\hat Q(v,z)$ is conjugate to $az$.  Hence $\hat Q(\a\cdot v, z)= a^{-1}\hat Q(v,z) a$, which implies that  
\beq\label{pn}
(\hat Q^j(\a\cdot v, z))_+= a^{-1}(\hat Q^j(v, z))_+ a.
\eeq  Since $v$ is a solution of \eqref{iq}, we have 
$$[\p_x+ a z + P(v), \p_t+ (\hat Q^j(v, z))_+]=0.$$
Conjugate the above equation by $a^{-1}$ to get
\begin{align*}
&a[\p_x+ a z+ P(v), \p_t+ (\hat Q^j(v, z))_+]a^{-1}\\
&\quad = [\p_x + az + P(\a\cdot v), \p_t+ a^{-1}(\hat Q^j(v,z))_+ a], \quad {\rm by \, \eqref{pn},}\\
& \quad = [\p_x+ a z + P(\a\cdot v), \p_t+ (\hat Q^j(\a\cdot v, z))_+]=0.
\end{align*}
This proves that $\a\cdot v$ is a solution of \eqref{iq}.
\end{proof}

Next we use Theorem \ref{po} to show that Adler's BT arises naturally from the $Z_n$-action on the KW-flows.

\bcor\label{tg} Let $\Phi:\R^{1\times (n-1)}\to \ct_n$ be the linear isomorphism defined by $\Phi(v)=\diag(q_1, \ldots, q_n)$ in Theorem \ref{iva}, where $q_i= \sum_{i=1}^{n-1} \a^{k(i-1)}v_k$. Let $q=\diag(q_1, \ldots, q_n)$ be a solution of the $j$-th $n\times n$ mKdV flow, and $v=\Phi^{-1}(q)$ the corresponding solution of the $j$-th KW flow. Then  $\Phi(\a^{n-1}\cdot v)= \diag(q_n, q_1, \ldots, q_{n-1})$ and is a solution of the $j$-th $n\times n$ mKdV flow.
\ecor

\begin{proof}
By Theorem \ref{iva}, $v$ is a solution of the $j$-th KW flow. By Theorem \ref{po}, $\a\cdot v=(\a v_1, \ldots, \a^{n-1}v_{n-1})$ is a solution of the $j$-th KW flow.  Hence $\Phi(\a\cdot v)$ is a also a solution of the $j$-th $n \times n$ mKdV flow.  A simple computation implies that $\Phi(\a\cdot v)=\diag(q_2, \ldots, q_n, q_1)$.  So 
$$\Phi(\a^{n-1}\cdot v)= \diag(q_n, q_1, \ldots, q_{n-1})$$ is a solution of the $j$-th $n \times n$ mKdV flow. 
\end{proof}

This Corollary gives a proof of Adler's result: Suppose 
$$L=\p^n-\sum_{i=1}^{n-1} u_i\p^{(i-1)} = (\p-q_1)\cdots (\p- q_n)$$ such that $L$ is a solution of the $j$-th GD$_n$ flow \eqref{ri} and $q=\diag(q_1, \ldots, q_n)$ is a solution of the $j$-th $n\times n$ mKdV flow \eqref{im}. By Corollary \ref{tg}, $$\diag(q_n, q_1, \ldots, q_{n-1})$$ is  a solution of \eqref{im}. It follows from Theorem \ref{ivb} that $\ti L= (\p-q_n)(\p-q_1)\cdots (\p-q_{n-1})$ is a solution of \eqref{ri}. 

\bs
\section{Relation between Adler's BT and our BT} \label{pi}

In this section, we
 prove that if $L$ is a solution of the $j$-th GD$_n$ flow \eqref{ri}, then 
 $\ti L=(\p+h)^{-1}L (\p+h)$ is a solution of \eqref{ri} if and only if  $h$ is a solution of \eqref{btj}.  
 We also show that Adler's BT is our BT with parameter $k=0$. 

\bprop\label{tf}
Let $L$ be a solution of the $j$-th GD$_n$ flow \eqref{ri} and $h:\R^2\to \C$. Then $\ti L= (\p+h)^{-1}L (\p+h)$ is a $n$-th order differential operator and is a solution of \eqref{ri} if and only if 
\beq\label{bta}\bca h_x= r_n(u,h),\\ h_t= w_{n,j}(u,h),\eca
\eeq
 for some differential polynomial  $w_{n,j}(u,h)$, where $r_n(u,h)$ is as in Theorem \ref{mz}.
\eprop 

\begin{proof} 
It follows from Lemma \ref{ni} and the proof of Theorem \ref{mz} that $\ti L$ is in $\calD_n$ if and only if $h$ satisfies $h_x^{(n)}= r_n(u,h)$ and the coefficient $\ti u_i$ of $\p^{i-1}$ is given by \eqref{nm1}. So $\ti L$ is a solution of \eqref{ri} if and only if 
$$\ti L_t=[(\p+h)^{-1}(L^{\frac{j}{n}})_+(\p+h)-(\p+h)^{-1}h_t, \ti L].$$
Note that $\ti L^{j/n}= (\p+h)^{-1}L^{j/n} (\p+h)=\p^j+\cdots$, by comparing the highest power of $\p$, we have 
$$(\p+h)^{-1}(L^{\frac{j}{n}})_+(\p+h)-(\p+h)^{-1}h_t=(\ti L^{\frac{j}{n}})_+.$$ 
Hence we obtain
$$h_t=-(L^{\frac{j}{n}})_-(\p+h)+(\p+h)(\ti L^{\frac{j}{n}})_- .$$
Let $p_{-1}$, $\ti p_{-1}$ denote the coefficient of $\p^{-1}$ of $L^{j/n}$ and $\ti L^{j/n}$ respectively. 
 Since coefficients of $(L^{j/n})_+$ and $(L^{\frac{j+n}{n}})_+$ are differential polynomials in $u$ and $\ti u$ respectively, $p_{-1}$ and $\ti p_{-1}$ are differential polynomials in $u$ and $h$. 
 So we have
\beq\label{te}
h_t=w_{n,j}(u,h):= -p_{-1}(u, h)+\ti p_{-1}(u, h).
\eeq
Hence we proved the proposition.
\end{proof}

\bthm\label{sg} Let $L=\p^n-\sum_{i=1}^{n-1} u_i \p^{i-1}$ be a solution of the $j$-th GD$_n$ flow \eqref{ri}, and $h\in C^\infty(\R^2, \C)$. Then the following are equivalent:
\ben
\item[(a)] $\ti L= (\p+h)^{-1}L (\p+h)$ is a solution of the $j$-th GD$_n$ flow \eqref{ri},
\item[(b)] $h$ is a solution of \eqref{bta},
\item[(c)] $h$ is a solution of \eqref{btj},
\item[(d)] $h$ is a solution of $({\rm BT})_{u,k}$ \eqref{btn} for some constant $k\in \C$. 
\een
\ethm 

\begin{proof} The equivalence of (a) and (b) is given by Proposition \ref{tf}, and the equivalence of (c) and (d) is given by Corollary \ref{sp}. 
  
  We prove (a) implies (c).  
Suppose $\ti L=(\p+h)^{-1}L (\p+h)$ is a solution of \eqref{ri}. Write $\ti L=\p^n-\sum_{i=1}^{n-1}\ti u_i\p^{i-1}$. By Theorem \ref{so}, both $u=\sum_{i=1}^{n-1} u_i e_{in}$ and $\ti u=\sum_{i=1}^{n-1}\ti u_i e_{in}$ are solutions of \eqref{dga}. Let $\ti E(x,t,\l)$ be a frame of the solution $\ti u$, and $\ti\phi=(\ti\phi_1, \ldots, \ti\phi_n)^t$ the first column of $\ti E$. It follows from Lemma \ref{ni} that $\{\ti\phi_1, \ldots, \ti\phi_n\}$ are solutions of $\ti L-\l =0$.
Since $L -\l=(\p+h)(\ti L-\l)(\p+h)^{-1}$, $\phi_i=(\p+h)\ti \phi_i$ is a solution of $L-\l=0$ for $1\leq i\leq n$. Let $E=((\phi_i)_x^{(j-1)})$, and $f= \ti E^{-1} E$. So $E=\ti E f$. A direct computation implies that the first column of $f$ is $(h,1,\ldots, 0)^t$ and $\det(f(x,t,\l))$ is of degree one in $\l$. Since $\ti E$ is a frame for $\ti u$, $\det(\ti E)$ is independent of $x,t$, is an analytic function $c(\l)$, and $c(\l)\not=0$ for all $\l\in \C$. So $\det(E(x,t,\l))\not=0$ for generic $(x,t,\l)$, i.e., $\{\phi_1, \ldots, \phi_n\}$ is a basis of $L-\l=0$ for generic $\l$.  Hence $E$ is a frame of $u$.  By Theorem \ref{mz}, $E=\ti E f_{u, h}$ and $h$ is a solution of \eqref{btj}.  

Next we prove (d) implies (a). Suppose $h$ is a solution of (BT)$_{u,k}$ \eqref{btn}. By Theorem \ref{ow} we have 
\ben
\item[(i)] $\ti u= h\ast u$ is a solution of \eqref{dga}, 
\item[(ii)] if $\ti E$ is a frame of $\ti u$, then $E= \ti Ef_{u,h}$ is a frame for $u$ for all $\l\not=k$.
\item[(iii)] the $i1$-th entries of $\phi_i$ $E$ and $\ti \phi_i$ $\ti E$ are related by $\phi_i= (\p+h) \ti\phi_i$. 
\een
 Assume $\l\not=k$.  Lemma \ref{ni} implies that $\{\phi_1, \ldots, \phi_n\}$ is a basis of $L-\l=0$. This is also a basis of $(\p+h)(\ti L-\l) (\p+h)^{-1}=0$,  where $\ti L=\p^n-\sum_{i=1}^{n-1} \ti u_i \p^{(i-1)}$. It follows from Lemma \ref{ni} that $L-\l=(\p+h)(\ti L-\l)(\p+h)^{-1}$.  Hence $L=(\p+h)\ti L(\p+h)^{-1}$.  Or equivalently, $\ti L=(\p+h)^{-1} L(\p+h)$.  
\end{proof}

Although both $\eta_{n,j}(u,h)$ and $w_{n,j}(u,h)$ can be computed, these two differential polynomials are not in explicit closed form; hence we do not know whether they are the same. But systems \eqref{bta} and \eqref{btj} have the same set of solutions.  

\bcor\label{ss}  Let $L=\p^n-\sum_{i=1}^{n-1} u_i \p^{i-1}$ be a solution of the $j$-th GD$_n$ flow \eqref{ri}, $u=\sum_{i=1}^{n-1} u_i e_{in}$, and $k\in \C$ a constant. Let $p_{j,i}(u,\l)$ denote the $i1$-th entry of $Z_j(u,k)$, where $Z_j(u,\l)$ is defined by \eqref{pp} for the $j$-th $\an1$-KdV flow. Then 
\beq\label{st}
\bca (L-k) \phi=0, \\ \phi_t= \sum_{i=1}^n p_{j,i}(u,k) \phi_x^{(i-1)},\eca
\eeq
is solvable for $\phi:\R^2\to \C$,   Moreover, if $\phi_1, \ldots, \phi_{n-1}$ are linearly independent solutions of \eqref{st}, then
$$h=(\ln (W(\phi_1, \cdots, \phi_{n-1})))_x,$$
is a solution of $( {\rm BT})_{u,k}$ \eqref{btn} and $\ti L=(\p+h)^{-1}L(\p+h)$ is a solution of \eqref{ri}, where $W(f_1, \ldots, f_{n-1})= \det((f_i)_x^{(j-1)})$ is the Wronskian. 
\ecor

\begin{proof} By Theorem \ref{so}, $u$ is a solution of \eqref{dga}. Let $g=(\phi, \phi_x, \ldots, \phi_x^{(n-1)})$. Then system \eqref{st} is equivalent to 
\beq\label{sta}\bca g^{-1}g_x= ke_{1n}+ b+ u, \\ g^{-1}g_t= Z_j(u,k).\eca\eeq
This is the Lax pair of $u$ with $\l=k$, so it is solvable.  If $\phi$ is a solution of \eqref{st}, then $g$ satisfies \eqref{sta}. Let $(v_1, \ldots, v_n)^t= g^{-1}e_n$. It follows from Theorem \ref{wb} that $h=-\frac{v_{n-1}}{v_n}$ is a solution of \eqref{btn} with parameter $k$.  Use Cramer's rule to get the formula of $h$ in terms of Wronskians. 
\end{proof}

For example, when $n=3$, equation \eqref{st} is
\beq\label{th3}
\bca \phi_{xxx} -u_2 \phi_x - u_1\phi= \l \phi, \\ \phi_t= -\frac{2}{3} u_2 \phi + \phi_{xx}.\eca
\eeq 

Note that $\ti L$ and $L$ in Adler's Theorem satisfies the condition $\ti L= (\p-q_n)^{-1} L(\p-q_n)$. So by Theorem \ref{sg}, $-q_n$ is a solution of \eqref{btni} for some $k \in \C$.  Next we show that $k=0$. In other words, Adler's BT is our BT with parameter $k=0$. 

\bthm\label{mw}
 Let $L=\p^n-\sum_{i-1}^{n-1}u_i\p^{(i-1)} =(\p-q_n)\cdots (\p-q_1)$ be a solution of the $j$-th GD$_n$ flow \eqref{ri} such that $q=\sum_{i=1}^n q_i e_{ii}$ is a solution of the $j$-th $n\times n$ mKdV flow.  Then $-q_n$ is a solution of $({\rm BT})_{u,0}$, i.e., \eqref{btn} with parameter $k=0$.
\ethm

\begin{proof}
  Let $K$ denote the parallel frame for the Lax pair of the solution $q$ of the $j$-th $n\times n$ mKdV flow at $\l=0$ with $K(0,0)= \I_n$, i.e., $K$ is the solution of 
$$\bca K^{-1}K_x= b+ q,\\ K^{-1}K_t= \pi_{bn}(Q_{j,0}(q)),\eca$$
where $Q_{j,0}(q)$ is defined in section \ref{km}, $b=\sum_{i=1}^{n-1} e_{i+1, i}$, and $\pi_{bn}$ is the projection of $sl(n,\C)$ onto $\calB_n^-$ along $\calN_n^+$.  Since both $b+q$ and $\pi_{bn}(Q_{j,0}(q))$ are lower triangular and $K(0,0)=\I_n$, we see that $K(x,t)\in B_n^-$, i.e. lower triangular, for all $x,t$.  Let $\phi=(\phi_1, \ldots, \phi_n)^t$ denote the first column of $P$. It follows from $K_x= K(b+q)$ that the $j$-th column of $K$ is 
$$(\p- q_{j-1})\cdots (\p- q_1) \phi,$$
and $L\phi=(\p-q_n)\cdots (\p-q_1)\phi=0$. Since $K$ is lower triangular, we obtain 
\beq\label{rj}
(\p-q_{i})\cdots (\p-q_1) \phi_i=0,
\eeq
for $1 \leq i\leq n-1$.  

Let $\D(\cdot, t):= \G(q(\cdot, t))$, where $\G$ is given in Definition \ref{sm}.  By Theorem \ref{ivb}, $g= K\D^{-1}$ is a frame of the Lax pair of  the solution $u$ of \eqref{dga} at $\l=0$.  Note that $\D$ is in $N_n^+$, so the first column of $g$ is also $\phi$.  Use $g^{-1}g_x= b+u$ to obtain that $g=(\phi, \phi_x, \ldots, \phi_x^{(n-1)})$. 

Let
$$R:=(\p-q_{n-1}) \cdots (\p- q_1)=\p^{n-1} -\sum_{i=1}^{n-1} \xi_i \p^{i-1}.$$
By \eqref{rj},  $R \phi_j=0$ for all $1\leq j\leq n-1$. So we have
$$\sum_{i=1}^{n-1}\xi_i \phi_j^{(i-1)}= \phi_j^{(n-1)}, \quad 1\leq j\leq n-1.$$
By Cramer's rule, 
\beq\label{nn}
\xi_{n-1}= \frac{\det(\eta, \ldots, \eta_x^{(n-4)}, \eta_x^{(n-3)}, \eta_x^{(n-1)})}{\det(\eta, \eta_x, \ldots, \eta_x^{(n-2)})},
\eeq
where $\eta=(\phi_1, \ldots, \phi_{n-1})^t$.  Recall that $g=(\phi, \ldots, \phi_x^{(n-1)})$.  Let $g^{ij}$ denote the $ij$-th entry of $g^{-1}$.  Then the numerator and denominator of the right hand side of \eqref{nn} are $-\det(g) g^{n-1,n}$ and $\det(g)g^{nn}$ respectively. So $\xi_n=- \frac{g^{n-1,n}}{g^{nn}}$. By Theorem \ref{wb}, $\xi_{n-1}$ is a solution of \eqref{btni} with $k=0$.  
Compare the coefficient of $\p^{n-1}$ of $L=(\p-q_n)R$ to obtain $q_n= -\xi_{n-1}$.  Hence $-q_n$ is a solution of \eqref{btni} with $k=0$.
\end{proof}

\bs

\end{document}